\documentclass[%
 reprint,
showkeys,
 amsmath,amssymb,
 aps,
pre,
]{revtex4-2}

\usepackage{graphicx}					
\usepackage{tabularx}
\usepackage{dcolumn}					
\usepackage{bm}							
\usepackage{hyperref}					
\usepackage{units}							
\usepackage{makecell} 					
\usepackage{rotating}					
\usepackage{accents}		
\newcommand*{\dt}[1]{%
  \accentset{\mbox{\large\bfseries .}}{#1}}

\makeatletter 
\newcommand{\vast}{\bBigg@{4}}
\newcommand{\Vast}{\bBigg@{5}}
\makeatother

\begin{document}


\title{Viscous--Viscoelastic Correspondence Principle for Brownian Motion}

\author{Nicos Makris}
 \email{nmakris@smu.edu}
 \altaffiliation[Also at ]{Office of Theoretical and Applied \\Mechanics, Academy of Athens, 10679, Greece}
\affiliation{%
 Dept. of Civil and Environmental Engineering, Southern Methodist University, Dallas, Texas, 75276
}%

\date{\today}

\begin{abstract}
Motivated from the classical expressions of the mean squared displacement and the velocity autocorrelation function of Brownian particles suspended either in a Newtonian viscous fluid or trapped in a harmonic potential, we show that for all time-scales the mean squared displacement of Brownian microspheres with mass $m$ and radius $R$ suspended in any linear, isotropic viscoelastic material is identical to the creep compliance of a linear mechanical network that is a parallel connection of the linear viscoelastic material with an inerter with distributed inertance, $m_R=$ {\normalsize $\frac{m}{\text{6}\pi R}$}. The synthesis of this mechanical network leads to the statement of a viscous--viscoelastic correspondence principle for Brownian motion which simplifies appreciably the calculations of the mean squared displacement and the velocity autocorrelation function of Brownian particles suspended in viscoelastic materials where inertia effects are non-negligible at longer time-scales. The viscous--viscoelastic correspondence principle established in this paper by introducing the concept of the inerter is equivalent to the viscous--viscoelastic analogy adopted by \citet{MasonWeitz1995}.\end{abstract}

\maketitle


\section{Introduction}
In microrheology the macroscopic frequency and time--response functions of linear viscoelastic materials are extracted by monitoring the thermally-driven Brownian motion of probe microspheres suspended within the viscoelastic material and subjected to random forces. The thermal fluctuations of the suspended probe particles have been monitored either with diffusing wave spectroscopy \citep{WeitzPine1993, MasonWeitz1995, MasonGangWeitz1997, PalmerXuWirtz1998, XuViasnoffWirtz1998} or with laser interferometry \citep{GittesSchnurrOlmstedMacKintoshSchmidt1997, SchnurrGittesMacKintoshSchmidt1997, MasonGanesanVanZantenWirtzKuo1997, LiKheifetsMedellinRaizen2010, HuangChavezTauteLukicJeneyRaizenFlorin2011} with nanometer spatial resolution and sub-microsecond temporal resolution. In their seminal paper, \citet{MasonWeitz1995} derived an expression that relates the complex dynamic modulus, $\mathcal{G}_{\text{VE}}(s)$, of the viscoelastic material surrounding the probe microspheres with the Laplace transform of the mean squared displacement, $\left\langle \Delta^{\text{2}}(s) \right\rangle$, of the suspended microspheres with mass $m$ and radius $R$. While the original \citet{MasonWeitz1995} expression includes an inertia term; it was noted that it is negligible in the applications of interest except at very high frequencies. Accordingly, subsequent publications \citep{MasonGanesanVanZantenWirtzKuo1997, MasonGangWeitz1997, Mason2000, GardelValentineWeitz2005, SquiresMason2010} neglect the inertia term $ms^{\text{2}}$ and concentrate on the following relation in the frequency domain
\begin{equation}\label{eq:Eq01}
\mathcal{G}_{\text{VE}}(s)=\frac{N K_B T}{\text{3}\pi R s \left\langle \Delta^{\text{2}}(s) \right\rangle}
\end{equation}
where $s$ is the Laplace variable, $N\in \left\lbrace \text{1, 2, 3}  \right\rbrace$ is the number of spatial dimensions, $K_B$ is Boltzmann's constant and $T$ is the equilibrium temperature. By noting that the Laplace transform of the creep compliance, $J_{\text{VE}}(t)$ (strain history due to a unit step--stress) is $\mathcal{L}\left\lbrace J_{\text{VE}}(t) \right\rbrace=\displaystyle\int_{\text{0}}^{\infty}J_{\text{VE}}(t)e^{-st}\mathrm{d}t=$ {\large $\frac{\text{1}}{s\mathcal{G}_{\text{VE}}(s)}$}, \citet{PalmerXuWirtz1998}, and \citet{XuViasnoffWirtz1998} presented the following relation in the time domain
\begin{equation}\label{eq:Eq02}
\left\langle \Delta r^{\text{2}}(t) \right\rangle= \frac{N K_B T}{\text{3}\pi R} J_{\text{VE}}(t)
\end{equation}
which relates the mean squared displacement $\left\langle \Delta^{\text{2}}(t) \right\rangle$ of the suspended Brownian particles with the creep compliance $J_{\text{VE}}(t)$ of the viscoelastic material within which the Brownian particles are suspended. \citet{PalmerXuWirtz1998} and in a subsequent paper \citet{XuViasnoffWirtz1998} acknowledged that Eq. \eqref{eq:Eq02} neglects inertia effects and indicated that for the limiting case of Brownian motion of microspheres in a Newtonian fluid with viscosity $\eta$, Eq. \eqref{eq:Eq02} predicts the classical \citet{Einstein1905} result $\left\langle \Delta r^{\text{2}}(t) \right\rangle=$ {\large $\frac{N K_B T}{\text{3}\pi R} \frac{\text{1}}{\eta}$}$t$, since the creep compliance of the Newtonian viscous fluid is $ J_{\text{VE}}(t)=$ {\large $\frac{\text{1}}{\eta}$}$t$ \citep{BirdArmstrongHassager1987}. Clearly, Eq. \eqref{eq:Eq02} yields the classical \citet{Einstein1905} result, valid for longer time scales, because the inertia term $ms^{\text{2}}$ in the original \citet{MasonWeitz1995} equation was neglected. At short time scales \citep{LiKheifetsMedellinRaizen2010, HuangChavezTauteLukicJeneyRaizenFlorin2011, LiRaizen2013}, when $t<$ {\large $\frac{m}{\text{6}\pi R \eta}$} $=\tau$, the Brownian motion of suspended particles is influenced by the inertia of the particle and the surrounding fluid and \citeauthor{Einstein1905}'s ``long-term'' result was extended for the entire time-regime by \citet{UhlenbeckOrnstein1930}
\begin{equation}\label{eq:Eq03}
\left\langle \Delta r^{\text{2}}(t) \right\rangle=\frac{N K_B T}{\text{3} \pi R} \frac{\text{1}}{\eta} \left[ t-\tau \left( \text{1}-e^{-\frac{t}{\tau}} \right) \right]
\end{equation}
where $\tau=$ {\large $\frac{m}{\text{6}\pi R \eta}$}  is the dissipation time-scale of the perpetual fluctuation--dissipation process.

The quantity in brackets on the right-hand side of Eq. \eqref{eq:Eq03} which accounts for inertia effects is a time-response function that is different from the creep compliance of the surrounding Newtonian, viscous fluid, $J_{\text{VE}}(t)=$ {\large $\frac{\text{1}}{\eta}$}$t$. In the following section we show that the creep compliance $J(t)=$ {\large $\frac{\text{1}}{\eta}$} $ \left[ t-\tau \left( \text{1}-e^{-\frac{t}{\tau}} \right) \right]$ is the creep compliance (retardation function) of a linear network where a dashpot with viscosity $\eta$ is connected in parallel with an inerter with distributed inertance $m_R=$ {\large $\frac{m}{\text{6}\pi R}$}.

\section{The Inertoviscous Fluid}\label{sec:Sec02}
The equation of motion of Brownian microspheres with radius $R$ and mass $m$, suspended in a Newtonian, viscous fluid with viscosity $\eta$ when subjected to the random forces $f_R(t)$ that originate from the collisions of the fluid molecules on the Brownian particles (microspheres) is described by the Langevin equation \citep{Waigh2005, Attard2012, CoffeyKalmykov2017}. 
\begin{equation}\label{eq:Eq04}
m\frac{\mathrm{d}v(t)}{\mathrm{d}t}=-\zeta v(t)+f_R(t)
\end{equation}
where $v(t)=$ {\large $\frac{\mathrm{d}r(t)}{\mathrm{d}t}$} is the particle velocity and $\zeta v(t)$ is a viscous drag force proportional to the velocity of the Brownian particle. For a memoryless viscous fluid with viscosity $\eta$, the drag coefficient is given by Stokes' law; $\zeta=\text{6}\pi R \eta$ \citep{LandauLifshitz1959}. Upon dividing with the mass $m$ of the particle, Eq. \eqref{eq:Eq04} assumes the form 
\begin{equation}\label{eq:Eq05}
\frac{\mathrm{d}v(t)}{\mathrm{d}t}+\frac{\text{1}}{\tau}v(t)=\frac{f_R(t)}{m}
\end{equation}
where $\tau=$ {\large $\frac{m}{\text{6}\pi R\eta}$}. The random excitation, $f_R(t)$, has a zero average value over time, $\left\langle f_R(t) \right\rangle = \text{0}$; while for the memoryless viscous fluid that only dissipates energy, the correlation function contracts to a Dirac delta function \citep{Lighthill1958}.
\begin{equation}\label{eq:Eq06}
\left\langle f_R(t_{\text{1}})f_R(t_{\text{2}})  \right\rangle = A \delta(t_{\text{1}}-t_{\text{2}})
\end{equation} 
with $A$ being a constant that expresses the strength of the random forces. 

\begin{figure}[t!]
\centering
\includegraphics[width=0.8\linewidth, angle=0]{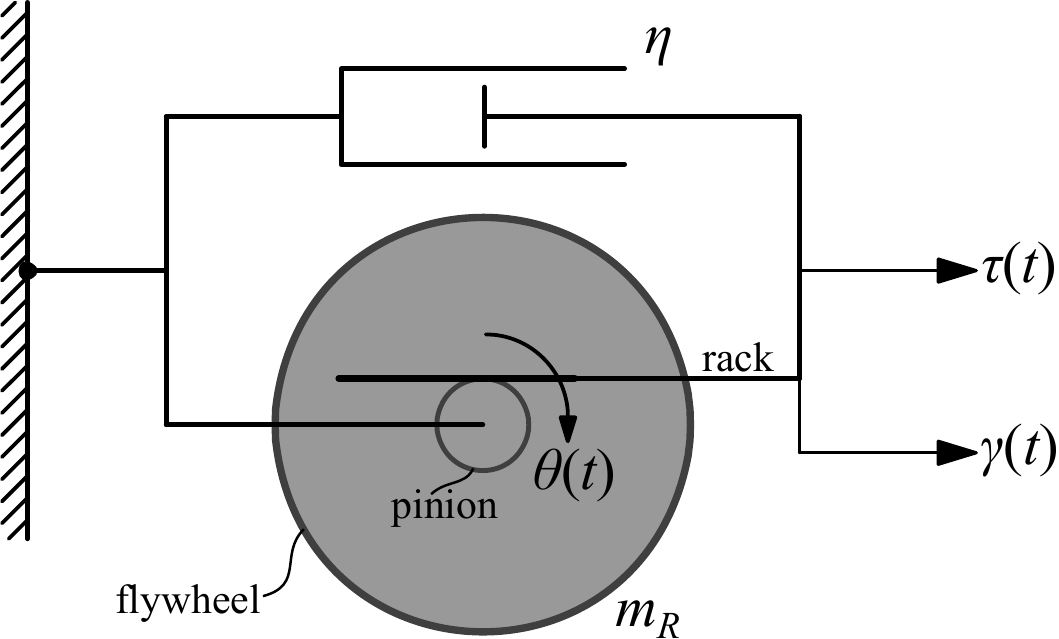}
\caption{The inertoviscous fluid which is a parallel connection of an inerter with distributed inertance $m_R$ with units $\left[\text{M}\right]\left[\text{L}\right]^{-\text{1}}$ and a dashpot with viscosity $\eta$ with units $\left[\text{M}\right]\left[\text{L}\right]^{-\text{1}}\left[\text{T}\right]^{-\text{1}}$. In analogy with the traditional schematic of a dashpot that is a hydraulic piston, the distributed inerter is depicted schematically with a rack--pinion--flywheel system \citep{Smith2002}.}
\label{fig:Fig01}
\end{figure}

Given the random nature of the excitation force, $f_R(t)$, the Langevin equation \eqref{eq:Eq05} can be integrated in terms of ensemble averages in association with Eq. \eqref{eq:Eq06} and the mean squared displacement of particles suspended in a viscous fluid is offered by Eq. \eqref{eq:Eq03} \citep{UhlenbeckOrnstein1930, WangUhlenbeck1945}. The synthesis of the proposed macroscopic mechanical network is suggested from the left-hand side of the Langevin equation \eqref{eq:Eq05}, which consists of an inertia term and a viscous term acting in parallel to balance the random force $f_R(t)$. Accordingly, we examine the frequency and time-response functions of the inertoviscous fluid shown in Fig. \ref{fig:Fig01} which is a parallel connection of a dashpot with viscosity $\eta$, with an inerter with distributed inertance $m_R$ \citep{Makris2017, Makris2018}.

An ``inerter'' is a linear mechanical element which at the force--displacement level, the output force is proportional only to the relative acceleration of its end-nodes (terminals) \citep{Smith2002, PapageorgiouSmith2005, Makris2017, Makris2018} and complements the set of the three elementary mechanical elements --- the other two elements being the traditional elastic spring and the viscous dashpot. In a force--current/ velocity--voltage analogy, the inerter is the mechanical analogue of the electric capacitor and its constant of proportionality is the ``inertance'' with units of mass $\left[\text{M}\right]$. For instance, a driving spinning top (with a steep lead angle) is a physical realization of an inerter, since the driving force is only proportional to the relative acceleration of its terminals. At the stress-strain level, the constant of proportionality of the inerter shown schematically in Fig. \ref{fig:Fig01} is the distributed inertance, $m_R$ with units $\left[\text{M}\right]\left[\text{L}\right]^{-\text{1}}$ (say \textit{Pa$\cdot$sec}$^{\text{2}}$).

Given the parallel connection of the dashpot and the inerter shown in Fig. \ref{fig:Fig01}, the constitutive law of the inertoviscous fluid is
\begin{equation}\label{eq:Eq07}
\tau (t)= \eta \frac{\mathrm{d}\gamma (t)}{\mathrm{d}t}+m_R \frac{\mathrm{d}^{\text{2}}\gamma(t)}{\mathrm{d}t^{\text{2}}}
\end{equation}
The Laplace transform of Eq. \eqref{eq:Eq07} gives
\begin{equation}\label{eq:Eq08}
\tau (s)=\mathcal{G}(s)\mathcal{\gamma}(s)=\left( \eta s + m_R s^{\text{2}} \right)\mathcal{\gamma}(s)
\end{equation}
where $\mathcal{G}(s)=\eta s + m_R s^{\text{2}}$ is the complex dynamic modulus. The complex dynamic compliance of the inertoviscous fluid given by Eq. \eqref{eq:Eq07} is
\begin{equation}\label{eq:Eq09}
\mathcal{J}(s)=\frac{\text{1}}{\mathcal{G}(s)}=\frac{\text{1}}{\eta s + m_R s^{\text{2}}} = \frac{\text{1}}{\eta}\left( \frac{\text{1}}{s} - \frac{\text{1}}{s+\frac{\text{1}}{\tau}} \right)
\end{equation}
where $\tau=$ {\large $\frac{m_R}{\eta}$} is the dissipation time --- that is a time-scale needed for the kinetic energy stored in the inerter with distributed inertance, $m_R$, to be dissipated by the dashpot with viscosity, $\eta$. For the inertoviscous fluid described by Eq. \eqref{eq:Eq07} and the Brownian particle--Newtonian fluid system described by the Langevin equation \eqref{eq:Eq05} to have the same dissipation time, $\tau=$ {\large $\frac{m_R}{\eta}$} = {\large $\frac{m}{\text{6}\pi R\eta}$}, the distributed inertance of the inerter needs to assume the value $m_R=$ {\large $\frac{m}{\text{6}\pi R}$}. 

The Laplace transform of the creep compliance $J(t)$ is the complex creep function $\mathcal{C}(s)=\mathcal{L}\left\lbrace J(t) \right\rbrace=\displaystyle\int_{\text{0}}^{\infty}J(t)e^{-st}\mathrm{d}t=$ {\large $\frac{\mathcal{J}(s)}{s}$} \citep{PalmerXuWirtz1998, EvansTassieriAuhlWaigh2009, Makris2019}. Upon dividing Eq. \eqref{eq:Eq09} with the Laplace variable, $s$
\begin{align}\label{eq:Eq10}
\mathcal{C}(s)=\frac{\mathcal{J}(s)}{s}=&\frac{\text{1}}{\eta}\left[ \frac{\text{1}}{s^{\text{2}}} - \frac{\text{1}}{s\left(s+\frac{\text{1}}{\tau}\right)}\right]=\\ \nonumber
& \frac{\text{1}}{\eta} \left[ \frac{\text{1}}{s^{\text{2}}} - \tau \left( \frac{\text{1}}{s} - \frac{\text{1}}{s+\frac{\text{1}}{\tau}} \right) \right]
\end{align}
Inverse Laplace transform of Eq. \eqref{eq:Eq10} offers the creep compliance of the inertoviscous fluid 
\begin{equation}\label{eq:Eq11}
J(t)=\mathcal{L}^{-\text{1}} \left\lbrace \mathcal{C}(s) \right\rbrace = \frac{\text{1}}{\eta} \left[ t - \tau\left( \text{1} - e^{-\frac{t}{\tau}} \right) \right]
\end{equation}
By comparing the results of Eqs. \eqref{eq:Eq03} and \eqref{eq:Eq11}, the mean squared displacement, $\left\langle \Delta r^{\text{2}}(t) \right\rangle$, of Brownian microspheres with radius $R$ suspended in a memoryless viscous fluid with viscosity $\eta$ is given by
\begin{equation}\label{eq:Eq12}
\left\langle \Delta r^{\text{2}}(t) \right\rangle = \frac{NK_B T}{\text{3}\pi R}J(t)
\end{equation}
The creep compliance, $J(t)$, appearing in the right-hand-side of Eq. \eqref{eq:Eq12} is not the creep compliance, $J_{\text{VE}}(t)=$ {\large $\frac{\text{1}}{\eta}$}$t$ of the viscous fluid within which the microspheres are suspended as is approximated by Eq. \eqref{eq:Eq02} proposed by \cite{PalmerXuWirtz1998, XuViasnoffWirtz1998}; rather it is the creep compliance of the inertoviscous fluid schematically shown in Fig. \ref{fig:Fig01} and expressed by Eq. \eqref{eq:Eq11}. The inerter connected in parallel with the viscous dashpot accounts for the inertia effects on the Brownian microsphere--Newtonian fluid system which prevail at short time scales \citep{UhlenbeckOrnstein1930, WangUhlenbeck1945, LiRaizen2013}. As time increases, Eq. \eqref{eq:Eq12} reduces to Eq. \eqref{eq:Eq02}, leading to the classical result $\left\langle \Delta r^{\text{2}}(t) \right\rangle=$ {\large $\frac{NK_B T}{\text{3}\pi R}\frac{\text{1}}{\eta}$}$t$ derived by \citet{Einstein1905}, which involves only the viscosity of the surrounding Newtonian fluid.

\section{Brownian Motion of a Particle in a Harmonic Trap}
The Brownian motion of a particle in a harmonic trap when excited by a random force, $f_R(t)$, has been studied by \citet{UhlenbeckOrnstein1930} and \citet{WangUhlenbeck1945}. The equation of motion of a microsphere with mass $m$ in a harmonic trap with viscous damping subjected to a random excitation force $f_R(t)$ is
\begin{equation}\label{eq:Eq13}
m\frac{\mathrm{d}^{\text{2}}r(t)}{\mathrm{d}t^{\text{2}}} + \zeta \frac{\mathrm{d}r(t)}{\mathrm{d}t} + kr(t) = f_R(t)
\end{equation}
where $r(t)$ is the particle displacement, $\zeta${\large $\frac{\mathrm{d}r(t)}{\mathrm{d}t}$} is a viscous drag force and $kr(t)$ is a linear restoring force proportional to the displacement of the Brownian particle, $r(t)$. Upon dividing with the mass of the particle, $m$, Eq. \eqref{eq:Eq13} gives
\begin{equation}\label{eq:Eq14}
\frac{\mathrm{d}^{\text{2}}r(t)}{\mathrm{d}t^{\text{2}}}  + \frac{\text{1}}{\tau}\frac{\mathrm{d}r(t)}{\mathrm{d}t} + \omega_{\text{0}}^{\text{2}}r(t)=\frac{f_R(t)}{m}
\end{equation}
where $\tau=$ {\large $\frac{m}{\zeta}$} = {\large $\frac{m}{\text{6}\pi R\eta}$} is the dissipation time and $\omega_{\text{0}}=$ {\large $\sqrt{\frac{k}{m}}$} is the undamped natural angular frequency of the trapped particle. For $\omega_{\text{0}}\tau >$ {\large $\frac{\text{1}}{\text{2}}$}, the system described by Eq. \eqref{eq:Eq14} is underdamped; for $\omega_{\text{0}}\tau =$ {\large $\frac{\text{1}}{\text{2}}$} the system is critically damped; while for $\omega_{\text{0}}\tau <$ {\large $\frac{\text{1}}{\text{2}}$}, the system is overdamped.

The mean squared displacement of a Brownian particle in a harmonic trap has been evaluated by \citet{WangUhlenbeck1945} after computing the velocity autocorrelation function of the random process described by Eq. \eqref{eq:Eq14}. For the underdamped case {\large $\big($}$\omega_{\text{0}}\tau >$ {\large $\frac{\text{1}}{\text{2}}\big)$}, 
\begin{align}\label{eq:Eq15}
\left\langle \Delta r^{\text{2}}(t) \right\rangle = & \frac{\text{2} N K_B T}{m\omega_{\text{0}}^{\text{2}}} \times \\ \nonumber
& \left[ \text{1} - e^{-\frac{t}{\text{2}\tau}}\left( \cos(\omega_D t) + \frac{\text{1}}{\text{2}\omega_D \tau} \sin(\omega_D t) \right) \right]
\end{align}
where $N \in \left\lbrace \text{1, 2, 3} \right\rbrace$ is the number of spatial dimensions, $K_B$ is the Boltzman constant, $T$ is the equilibrium temperature of the medium--particle system and $\omega_D = \omega_{\text{0}}${\large $\sqrt{\text{{\normalsize 1 $-$}}\left( \frac{\text{1}}{\text{2}\omega_{\text{0}}\tau} \right)^{\text{{\scriptsize 2}}}}$} is the damped angular frequency of the trapped particle.

\begin{figure}[b!]
\centering
\includegraphics[width=0.8\linewidth, angle=0]{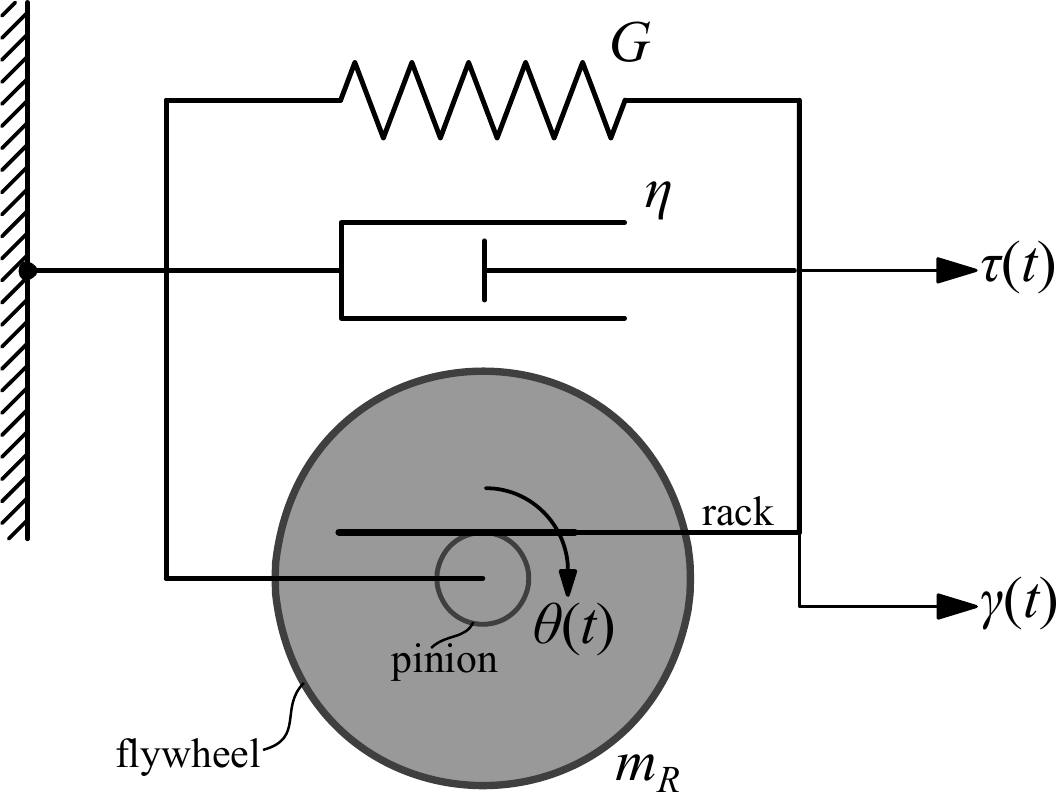}
\caption{The inertoviscoelastic solid which is a parallel connection of an inerter with distributed inertance $m_R$, a dashpot with viscosity $\eta$ and a linear spring with elastic shear modulus $G$.}
\label{fig:Fig02}
\end{figure}

The quantity in brackets of the right-hand side of Eq. \eqref{eq:Eq15} is a time--response function and by following the reasoning that introduced the inertoviscous fluid in Section \ref{sec:Sec02}, we are in search of a solid-like mechanical network that its creep compliance, $J(t)$, is of the form of the right-hand side of Eq. \eqref{eq:Eq15}. Again, the synthesis of the linear mechanical network is suggested from the left-hand side of the Langevin equation \eqref{eq:Eq14} which consists of an inertia term, a viscous term and a restoring term acting in parallel to balance the random excitation $f_R(t)$. Accordingly, we examine the frequency-- and time--response functions of the inertoviscoelastic solid shown in Fig. \ref{fig:Fig02} which is a parallel connection of a spring with elastic shear modulus $G$, a dashpot with viscosity $\eta$ and an inerter with distributed inertance $m_R$ \citep{Makris2018}. Given the parallel connection of the three elementary mechanical idealizations shown in Fig. \ref{fig:Fig02}, the constitutive law of the inertoviscoelastic solid is
\begin{equation}\label{eq:Eq16}
\tau(t)=G\gamma(t)+\eta\frac{\mathrm{d}\gamma(t)}{\mathrm{d}t}+m_R\frac{\mathrm{d}^{\text{2}}\gamma(t)}{\mathrm{d}t^{\text{2}}}
\end{equation}
The Laplace transform of Eq. \eqref{eq:Eq16} is
\begin{equation}\label{eq:Eq17}
\tau(s)=\mathcal{G}(s)\gamma(s)=\left( G + \eta s + m_R s^{\text{2}} \right)\gamma(s)
\end{equation}
where $\mathcal{G}(s)=G + \eta s + m_R s^{\text{2}}$ is the complex dynamic modulus; while the complex dynamic compliance is
\begin{align}\label{eq:Eq18}
\mathcal{J}(s)=\frac{\text{1}}{\mathcal{G}(s)} = & \frac{\text{1}}{G + \eta s + m_R s^{\text{2}}} = \\ \nonumber 
&\frac{\text{1}}{m_R} \frac{\text{1}}{\left( s + \frac{\text{1}}{\text{2}\tau} \right)^{\text{2}} + \omega_R^{\text{2}} - \left(\frac{\text{1}}{\text{2}\tau} \right)^{\text{2}}}
\end{align}
where again $\tau=$ {\large $\frac{m_R}{\eta}$} = {\large $\frac{m}{\text{6}\pi R \eta}$} is the dissipation time and $\omega_R=$ {\large $\sqrt{\frac{G}{m_R}}$} is the undamped rotational angular frequency of the inertoviscoelastic solid. For the inertoviscoelastic solid described by Eq. \eqref{eq:Eq16} and the Brownian particle in a harmonic trap described by the Langevin equation \eqref{eq:Eq13} to have the same undamped natural frequency, $\omega_R=$ {\large $\sqrt{\frac{G}{m_R}}$} $=$ {\large $\sqrt{\frac{k}{m}}$} $=\omega_{\text{0}}$; the shear modulus needs to assume the value $G=$ {\large $\frac{m_R}{m}$}$k$. Given that $m_R=$ {\large $\frac{m}{\text{6}\pi R}$}, the elastic shear modulus of the inertoviscoelastic solid described by Eq. \eqref{eq:Eq16} is $G=$ {\large $\frac{k}{\text{6}\pi R}$}, where $k$ is the spring constant of the harmonic trap appearing in Eq. \eqref{eq:Eq13}.

The Laplace transform of the creep compliance, $J(t)$, is the complex creep function $\mathcal{C}(s)=\displaystyle\int_{\text{0}}^{\infty}J(t)e^{-st}\mathrm{d}t=$ {\large $\frac{\mathcal{J}(s)}{s}$}, and Eq. \eqref{eq:Eq18} yields
\begin{align}\label{eq:Eq19}
\mathcal{C}(s)=\frac{\mathcal{J}(s)}{s}= & \, \mathcal{L}\left\lbrace J(t) \right\rbrace = \\ \nonumber 
& \frac{\text{1}}{m_R} \frac{\text{1}}{s} \frac{\text{1}}{\left( s + \frac{\text{1}}{\text{2}\tau} \right)^{\text{2}} + \omega_R^{\text{2}} - \left(\frac{\text{1}}{\text{2}\tau} \right)^{\text{2}}}
\end{align} 
Inverse Laplace transform of Eq. \eqref{eq:Eq19} offers the creep compliance (retardation function) of the inertoviscoelastic solid described by Eq. \eqref{eq:Eq16} \citep{Erdelyi1954}
\begin{align}\label{eq:Eq20}
J(t)= & \frac{\text{6}\pi R}{m\omega_{\text{0}}^{\text{2}}} \times \\ \nonumber
& \left[ \text{1} - e^{-\frac{t}{\text{2}\tau}} \left( \cos(\omega_D t) + \frac{\text{1}}{\text{2}\tau\omega_D} \sin(\omega_D t) \right) \right]
\end{align}
where $\omega_D = \omega_R${\large $\sqrt{\text{{\normalsize 1 $-$}}\left( \frac{\text{1}}{\text{2}\omega_R\tau} \right)^{\text{{\scriptsize 2}}}}$} $= \omega_{\text{0}}${\large $\sqrt{\text{{\normalsize 1 $-$}}\left( \frac{\text{1}}{\text{2}\omega_{\text{0}}\tau} \right)^{\text{{\scriptsize 2}}}}$} is the damped angular frequency. In deriving Eq. \eqref{eq:Eq20} we used $m_R=$ {\large $\frac{m}{\text{6}\pi R}$}. By comparing the results of Eqs. \eqref{eq:Eq15} and \eqref{eq:Eq20}, the mean squared displacement $\left\langle \Delta r^{\text{2}}(t) \right\rangle$ of a Brownian microsphere with radius $R$ in a harmonic trap (Kelvin--Voigt solid) is given again by Eq. \eqref{eq:Eq12}, where now $J(t)$ is the creep compliance of the inertoviscoelastic solid given by Eq. \eqref{eq:Eq20}.

In a dimensionless form Eq. \eqref{eq:Eq15} or Eq. \eqref{eq:Eq20} which are for the underdamped case $\Big(\omega_{\text{0}}\tau >$ {\large $\frac{\text{1}}{\text{2}}$}$\Big)$ are expressed as
{\small{%
\begin{align}\label{eq:Eq21}
\frac{m\omega_{\text{0}}^{\text{2}}}{\text{2}N K_B T} & \left\langle \Delta r^{\text{2}}(t) \right\rangle = G J(t) = \\ \nonumber
& \text{1}-e^{-\frac{t}{\text{2}\tau}}\vast[ \cos\left( \omega_{\text{0}}\tau \sqrt{\text{1}-\left( \frac{\text{1}}{\text{2}\omega_{\text{0}} \tau} \right)^\text{2}} \frac{t}{\tau}\right) + \\
& \frac{\text{1}}{\text{2}\omega_{\text{0}}\tau\sqrt{\text{1}-\left( \frac{\text{1}}{\text{2}\omega_{\text{0}}\tau} \right)^\text{2}}} \sin\left( \omega_{\text{0}}\tau \sqrt{\text{1}-\left( \frac{\text{1}}{\text{2}\omega_{\text{0}} \tau} \right)^\text{2}} \frac{t}{\tau}\right) \vast] \nonumber
\end{align}}
\begin{figure}[b!]
\centering
\includegraphics[width=\linewidth, angle=0]{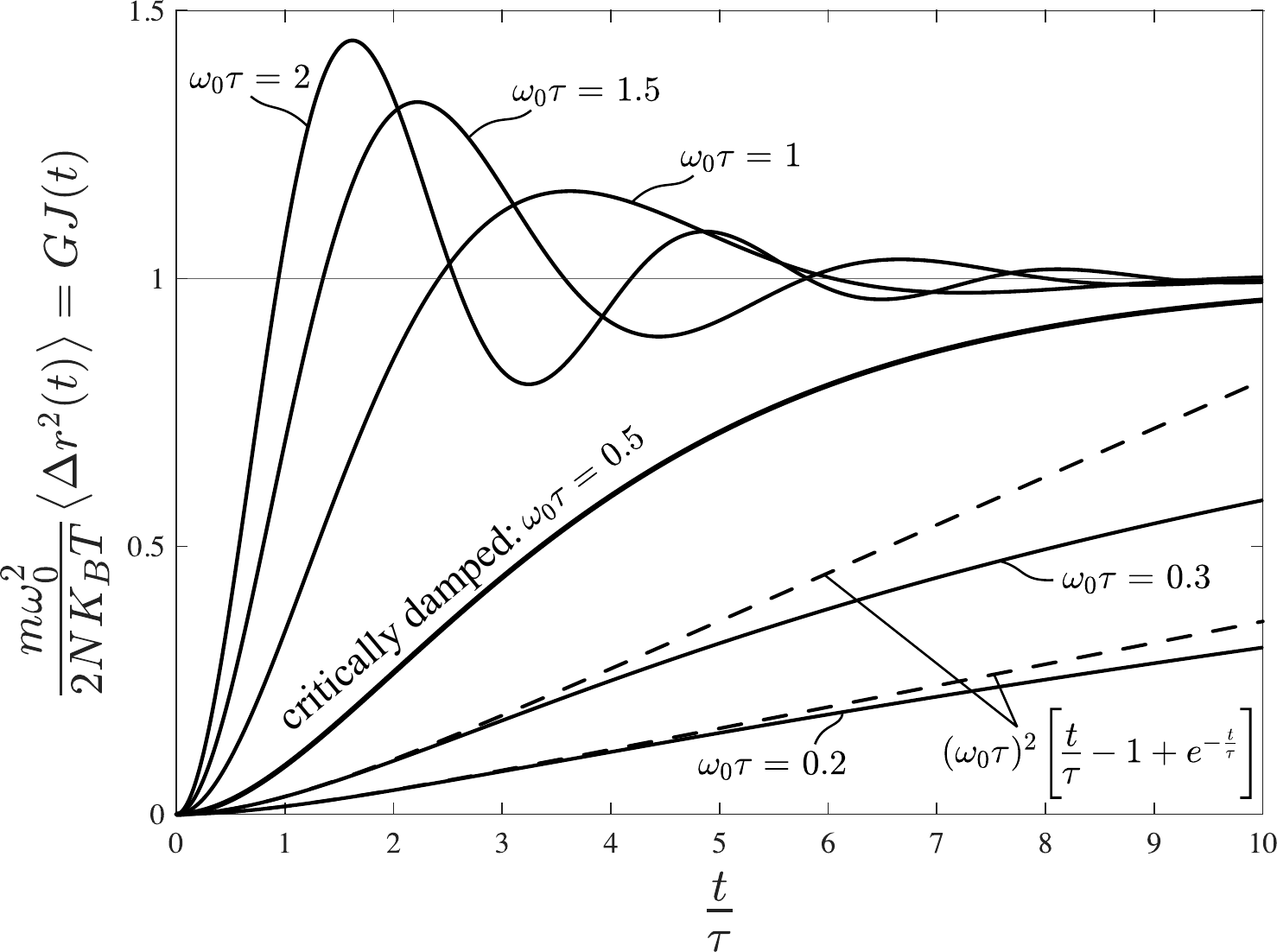}
\caption{Normalized mean squared displacement of a Brownian particle in a harmonic trap for the underdamped, $\omega_{\text{0}}\tau >$ {\large $\frac{\text{1}}{\text{2}}$}, critically damped, $\omega_{\text{0}}\tau =$ {\large $\frac{\text{1}}{\text{2}}$} and overdamped, $\omega_{\text{0}}\tau <$ {\large $\frac{\text{1}}{\text{2}}$} cases which is equal to $GJ(t)$ of the inertoviscoelastic solid shown in Fig. \ref{fig:Fig02}. For the overdamped cases (weak spring) at early times, the time--response functions of the Brownian particle in a harmonic trap coincide with the corresponding time--response functions of a Brownian particle in a viscous fluid with viscosity $\eta$.}
\label{fig:Fig03}
\end{figure}
\begin{figure*}[t!]
\centering
\includegraphics[width=0.75\linewidth, angle=0]{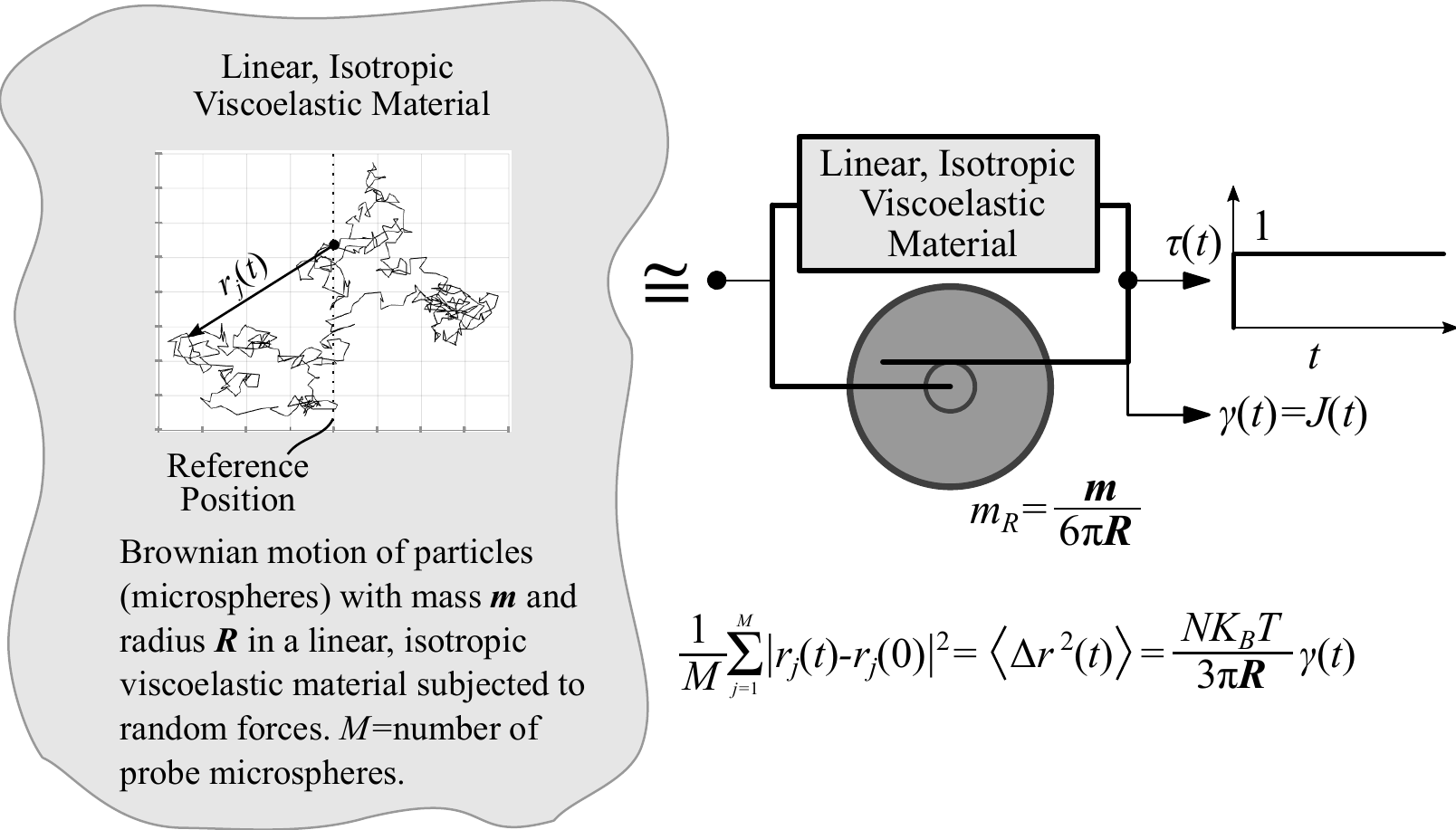}
\caption{Statement of the Correspondence Principle for Brownian motion. The mean squared displacement, $\left\langle \Delta r^{\text{2}}(t) \right\rangle$, of Brownian particles (microspheres) with mass $m$ and radius $R$ suspended in some linear, isotropic viscoelastic material when subjected to the random forces from the collisions of the molecules of the viscoelastic material, is identical to {\large $\frac{NK_B T}{\text{3}\pi R}$}$\gamma (t)$, where $\gamma (t)=J(t)$ is the strain due to a unit step--stress on a viscoelastic network that is a parallel connection of the linear viscoelastic material and an inerter with distributed inertance $m_r=$ {\large $\frac{m}{\text{6}\pi R}$} .}
\label{fig:Fig04}
\end{figure*}
For the overdamped case {\large $\big($}$\omega_{\text{0}}\tau <$ {\large $\frac{\text{1}}{\text{2}}\big)$}, the normalized mean squared displacement of a Brownian particle in a harmonic trap is 
{\small %
\begin{align}\label{eq:Eq22}
\frac{m\omega_{\text{0}}^{\text{2}}}{\text{2}N K_B T} & \left\langle \Delta r^{\text{2}}(t) \right\rangle = G J(t) = \\ \nonumber
& \text{1}-e^{-\frac{t}{\text{2}\tau}}\vast[ \cosh\left( \omega_{\text{0}}\tau \sqrt{\left( \frac{\text{1}}{\text{2}\omega_{\text{0}} \tau} \right)^\text{2}-\text{1}} \, \frac{t}{\tau}\right) + \\
& \frac{\text{1}}{\text{2}\omega_{\text{0}}\tau\sqrt{\left( \frac{\text{1}}{\text{2}\omega_{\text{0}}\tau} \right)^\text{2}-\text{1}}} \sinh\left( \omega_{\text{0}}\tau \sqrt{\left( \frac{\text{1}}{\text{2}\omega_{\text{0}} \tau} \right)^\text{2}-\text{1}} \, \frac{t}{\tau}\right) \vast] \nonumber
\end{align}}
For small values of the dimensionless product, $\omega_{\text{0}}\tau$ (weak spring), Eq. \eqref{eq:Eq22} at early times contracts to the solution for the Brownian motion of a particle in a Newtonian viscous fluid since the inertia and viscous terms dominate over the elastic term.
\begin{equation}\label{eq:Eq23}
\frac{m\omega_{\text{0}}^{\text{2}}}{\text{2}N K_B T} \left\langle \Delta r^{\text{2}}(t) \right\rangle = (\omega_{\text{0}}\tau)^{\text{2}}\left[ \frac{t}{\tau} -\text{1} + e^{-\frac{t}{\tau}} \right]
\end{equation}
Equation \eqref{eq:Eq23} is obtained after multiplying both sides of Eq. \eqref{eq:Eq03} with $\omega_{\text{0}}^{\text{2}}$ and replacing {\large $\frac{\text{1}}{\eta}$} with {\large $\frac{\text{6}\pi R \tau}{m}$}. Figure \ref{fig:Fig03} plots the normalized mean squared displacement given by Eqs. \eqref{eq:Eq21} and \eqref{eq:Eq22} as a function of the dimensionless time {\large $\frac{t}{\tau}$} for various values of $\omega_{\text{0}}\tau=$ {\large $\frac{\sqrt{k}\sqrt{m}}{\text{6}\pi R \eta}$} $=$ {\large $\frac{\sqrt{G}\sqrt{m_R}}{\eta}$} together with the results from Eq. \eqref{eq:Eq23} for values of $\omega_{\text{0}}\tau=$ 0.2 and 0.3.

\section{A Statement of a Correspondence Principle for Brownian Motion}\label{sec:Sec04}
The preceding analysis, leads invariably to Eq. \eqref{eq:Eq12} regardless whether the Brownian particles are suspended in a memoryless Newtonian fluid or a Kelvin--Voigt solid. The proposed correspondence principle for Brownian motion is partly motivated from the classical work by \citet{Lee1955} who established a correspondence principle between elastic and viscoelastic stress analysis. Given the linearity of viscoelastic constitutive laws, their time-dependence is removed by applying the Laplace transform and this enables the response analysis of a viscoelastic problem in terms of the solution of an associated elastic problem. With the elastic--viscoelastic correspondence principle established by \citet{Lee1955}, the extensive literature in the theory of elasticity can be utilized for solving problems in viscoelasticity \citep{Flugge1975, Pipkin1986, Tschoegl1989}.

Laplace transform of Eq. \eqref{eq:Eq12} gives
\begin{equation}\label{eq:Eq24}
\left\langle \Delta r^{\text{2}}(s) \right\rangle=\frac{N K_B T}{\text{3}\pi R}\frac{\text{1}}{s\mathcal{G}(s)}
\end{equation}
where $\mathcal{G}(s)=$ {\large $\frac{\text{1}}{\mathcal{J}(s)}$} is the complex dynamic modulus of the linear network shown on the right of Fig. \ref{fig:Fig04}, which is a parallel connection of the linear viscoelastic material with an inerter with distributed inertance $m_R=$ {\large $\frac{m}{\text{6}\pi R}$}. Given the parallel connection; the complex dynamic modulus of the linear network, $\mathcal{G}(s)$ is the superposition of the complex dynamic modulus of the linear viscoelastic material within which the Brownian particles are immersed, $\mathcal{G}_{\text{VE}}(s)$; and the complex dynamic modulus of the inerter $=m_R s^{\text{2}}=$ {\large $\frac{m}{\text{6}\pi R}$}$s^{\text{2}}$
\begin{equation}\label{eq:Eq25}
\mathcal{G}(s)=\mathcal{G}_{\text{VE}}(s)+\frac{m}{\text{6}\pi R}s^{\text{2}}
\end{equation}
Substitution of Eq. \eqref{eq:Eq25} into Eq. \eqref{eq:Eq24} gives
\begin{equation}\label{eq:Eq26}
\left\langle \Delta r^{\text{2}}(s) \right\rangle=\frac{N K_B T}{\text{3}\pi R} \frac{\text{1}}{s\left(\mathcal{G}_{\text{VE}}(s)+\frac{m}{\text{6}\pi R}s^{\text{2}}\right)}
\end{equation}
For the classical case of Brownian motion of particles in a memoryless, Newtonian viscous fluid with viscosity $\eta$, the Laplace transform of Eq. \eqref{eq:Eq03} gives 
\begin{equation}\label{eq:Eq27}
\left\langle \Delta r^{\text{2}}(s) \right\rangle=\frac{N K_B T}{\text{3}\pi R} \frac{\text{1}}{s\left(\eta s+\frac{m}{\text{6}\pi R}s^{\text{2}}\right)}
\end{equation}
Given that $\mathcal{G}_{\text{VE}}(s)=\eta_{\text{VE}}(s)s$, where $\eta_{\text{VE}}(s)$ is the complex dynamic viscosity of the linear, isotropic viscoelastic material, the direct analogy between Eqs. \eqref{eq:Eq26} and \eqref{eq:Eq27} leads to the following viscous--viscoelastic correspondence principle for Brownian motion: The mean squared displacement, $\left\langle \Delta r^{\text{2}}(t) \right\rangle$, of Brownian particles with mass $m$ and radius $R$ suspended in a linear, isotropic viscoelastic material (fluid or solid) due to the random forces from the collisions of the molecules of the viscoelastic material is identical to {\large $\frac{N K_B T}{\text{3}\pi R}$}$\gamma(t)$, where $\gamma(t)=J(t)$, is the strain due to a unit step--stress on a viscoelastic network that is a parallel connection of the linear visco-elastic material (within which the particles are suspended) and an inerter with distributed inertance $m_R=$ {\large $\frac{m}{\text{6}\pi R}$}.

The velocity autocorrelation function of the Brownian particles, $\left\langle v(\text{0})v(t) \right\rangle=\left\langle v(\xi)v(\xi+t) \right\rangle=\lim\limits_{T\to \infty}${\large $\frac{\text{1}}{T}$}$\displaystyle\int_{\text{0}}^{T}v(\xi)v(\xi+t) \mathrm{d}\xi$ is related with the mean squared displacement $\left\langle \Delta r^{\text{2}}(t) \right\rangle$ in the Laplace domain via the identity
\begin{equation}\label{eq:Eq28}
\mathcal{L}\left\lbrace \left\langle v(\text{0})v(t) \right\rangle \right\rbrace = \left\langle v(\text{0})v(s) \right\rangle = \frac{s^{\text{2}}}{\text{2}} \left\langle \Delta r^{\text{2}}(s) \right\rangle
\end{equation}
Substitution of Eq. \eqref{eq:Eq24} into Eq. \eqref{eq:Eq28} gives
\begin{equation}\label{eq:Eq29}
\left\langle v(\text{0})v(s) \right\rangle  = \frac{N K_B T}{\text{6}\pi R} \frac{s}{\mathcal{G}(s)}=\frac{N K_B T}{\text{6}\pi R}\frac{\text{1}}{\eta(s)}
\end{equation}
where $\eta(s)$ is the complex dynamic viscosity of the mechanical network shown on the right of Fig. \ref{fig:Fig04}. The inverse of the complex dynamic viscosity is known in rheology as the complex dynamic fluidity, $\phi(s)=$ {\large $\frac{\text{1}}{\eta(s)}$} $=$ {\large $\frac{\dt{\gamma}(s)}{\tau(s)}$} \citep{Giesekus1995, MakrisKampas2009} and relates a strain-rate output to a stress input. Accordingly, Eq. \eqref{eq:Eq29} is expressed as
\begin{equation}\label{eq:Eq30}
\left\langle v(\text{0})v(s) \right\rangle  = \frac{N K_B T}{\text{6}\pi R} \phi(s)
\end{equation}
In structural mechanics, the equivalent of the complex dynamic fluidity at the velocity--force level is known as the mechanical admittance or mobility \citep{HarrisCrede1976}. For the inertoviscous fluid shown in Fig. \ref{fig:Fig01}, the complex dynamic fluidity derives from Eq. \eqref{eq:Eq09}
\begin{equation}\label{eq:Eq31}
\phi(s)=\frac{s}{\mathcal{G}(s)}=\frac{\text{1}}{\eta(s)}=\frac{\text{1}}{\eta+m_R s}=\frac{\text{1}}{m_R} \, \frac{\text{1}}{s+\frac{\eta}{m_R}}
\end{equation}
and according to the proposed correspondence principle as expressed by Eq. \eqref{eq:Eq30}, the velocity autocorrelation function of Brownian particles suspended in a Newtonian viscous fluid is
\begin{equation}\label{eq:Eq32}
\left\langle v(\text{0})v(s) \right\rangle = \frac{N K_B T}{\text{6}\pi R} \frac{\text{1}}{m_R} \, \frac{\text{1}}{s+\frac{\eta}{m_R}}
\end{equation}
By using that the distributed inertance is $m_R=$ {\large $\frac{m}{\text{6}\pi R}$} and that {\large $\frac{\eta}{m_R}$} $=$ {\large $\frac{\text{6}\pi R \eta}{m}$} $=$ {\large $\frac{\text{1}}{\tau}$}, inverse Laplace transform of Eq. \eqref{eq:Eq32} gives
\begin{equation}\label{eq:Eq33}
\left\langle v(\text{0})v(t) \right\rangle = \frac{N K_B T}{m} \mathcal{L}^{-\text{1}}\left\lbrace \frac{\text{1}}{s+\frac{1}{\tau}} \right\rbrace = \frac{N K_B T}{m} e^{-\frac{t}{\tau}}
\end{equation}
which is the expected result for the velocity autocorrelation function of Brownian particles suspended in a memoryless viscous fluid \citep{UhlenbeckOrnstein1930}.

The inverse Laplace transform of the complex dynamic fluidity, $\mathcal{L}^{-\text{1}}\left\lbrace \phi(s) \right\rbrace=\psi(t)$ is the impulse strain-rate response function defined as the resulting strain-rate output at time $t$ due to an impulsive stress input at time $\xi$ $(\xi<t)$. Accordingly, an alternative statement of the visous--viscoelastic correspondence principle for Brownian motion uncovered in this study is that the velocity autocorrelation fucntion of Brownian microspheres with mass $m$ and radius $R$ suspended in some linear, isotropic viscoelastic material when subjected to the random forces from the collisions of the molecules of the viscoelastic material is
\begin{equation}\label{eq:Eq34}
\left\langle v(\text{0})v(t) \right\rangle = \frac{N K_B T}{\text{6}\pi R} \psi(t)
\end{equation}
where $\psi(t) = \mathcal{L}^{-\text{1}}\left\lbrace \phi(s) \right\rbrace$ is the impulse strain-rate response function of the mechanical network shown on the right of Fig. \ref{fig:Fig04}.

The impulse strain-rate response function of the inertoviscoelastic solid shown in Fig. \ref{fig:Fig02} was recently derived by \citet{Makris2018}
\begin{equation}\label{eq:Eq35}
\psi(t)=\frac{\text{1}}{m_R}\left[ \cos(\omega_D t) -\frac{\text{1}}{\text{2}} \frac{\text{1}}{\tau \omega_D}\sin(\omega_D t) \right]e^{-\frac{\text{1}}{\text{2}} \frac{t}{\tau}}
\end{equation}
where $\omega_D$ is the damped angular frequency as defined following Eq. \eqref{eq:Eq20} and $\tau=$ {\large $\frac{m_R}{\eta}$} $=$ {\large $\frac{m}{\text{6}\pi R \eta}$}. By using that $m_R=$ {\large $\frac{m}{\text{6}\pi R}$}, the substitution of Eq. \eqref{eq:Eq35} into Eq. \eqref{eq:Eq34} yields that the velocity autocorrelation of Brownian particles trapped in a harmonic potential is
\begin{align}\label{eq:Eq36}
\left\langle v(\text{0})v(t) \right\rangle = & \frac{N K_B T}{m} \times \\ \nonumber 
& \left[ \cos(\omega_D t) -\frac{\text{1}}{\text{2}} \frac{\text{1}}{\tau \omega_D}\sin(\omega_D t) \right]e^{-\frac{\text{1}}{\text{2}} \frac{t}{\tau}}
\end{align}
and the classical result derived by \citet{WangUhlenbeck1945} is recovered.

The main advantage of the proposed correspondence principle that is illustrated in Fig. \ref{fig:Fig04} is that it is valid for all time-scales; while the action of the random forces, $f_R$, on the Brownian particles is replaced with the action of a unit step--stress $\tau(t)=U(t-\text{0})$ on a linear viscoelastic network. This remarkable analogy reduces the mathematics involved for the solution of the stochastic generalized Langevin equation to the time--response analysis of a deterministic linear network.

From Eq. \eqref{eq:Eq26}, the complex dynamic modulus, $\mathcal{G}_{\text{VE}}(s)$, of the viscoelastic material within which the Brownian particles are suspended is given by
\begin{equation}\label{eq:Eq37}
\mathcal{G}_{\text{VE}}(s)=\frac{N K_B T}{\text{3}\pi R} \frac{\text{1}}{s\left\langle \Delta r^{\text{2}}(s) \right\rangle}-\frac{m}{\text{6}\pi R}s^{\text{2}}
\end{equation}
Eq. \eqref{eq:Eq37} has been derived by \citet{MasonWeitz1995} by following a different approach.

\section{Comparison with the Mason and Weitz Derivation}
\citet{MasonWeitz1995} employed dynamic light scattering to measure the mean squared displacement of probe particles, $\left\langle \Delta r^{\text{2}}(t) \right\rangle$, and related it to the complex dynamic modulus, $\mathcal{G}_{\text{VE}}(s)$, of the viscoelastic fluid within which the particles are suspended. Given the viscoelastic behavior of the complex fluid, the motion of a particle is described with the generalized Langevin equation \citep{VolkovVinogradov1984, RodriguezSalinas1988}
\begin{equation}\label{eq:Eq38}
m\frac{\mathrm{d}v(t)}{\mathrm{d}t}=-\int_{\text{0}}^{t}\zeta (t-\xi)v(\xi)\mathrm{d}\xi+f_R(t)
\end{equation} 
where $m$ is the particle mass, $v(t)$ is the particle velocity and $f_R(t)$ are random forces acting on the particle. The integral in Eq. \eqref{eq:Eq38} represents the drag force on the particle as it moves through the viscoelastic fluid and accounts for the fading memory of this drag due to the elasticity of the fluid. The elastic component of the fluid influences the temporal correlations of the stochastic forces acting on the particle; therefore, in this case Eq. \eqref{eq:Eq06} is replaced with
\begin{equation}\label{eq:Eq39}
\left\langle f_R(t) f_R(\text{0}) \right\rangle = K_B T \zeta(t-\text{0})
\end{equation}
where $\zeta(t-\text{0})$ is the relaxation kernel of the convolution appearing in the generalized Langevin equation \eqref{eq:Eq38}. \citet{MasonWeitz1995} calculated the mean squared displacement of the suspended particles in the Laplace domain by making the assumption that Stokes' result for the drag coefficient on a moving sphere in a memoryless viscous fluid, $\zeta = \text{6}\pi R\eta$ \citep{LandauLifshitz1959}, can be generalized to relate the complex dynamic viscosity of the viscoelastic material, $\eta_{\text{VE}}(s)=$ {\large $\frac{\mathcal{G}_{\text{VE}}(s)}{s}$} with the impedance of the Brownian particle--viscoelastic material system, $\mathcal{Z}(s)=\mathcal{L}\left\lbrace \zeta(t) \right\rbrace=\displaystyle\int_{\text{0}}^{t}\zeta(t)e^{-st}\mathrm{d}t$.
\begin{equation}\label{eq:Eq40}
\eta_{\text{VE}}(s)=\frac{\mathcal{G}_{\text{VE}}(s)}{s}=\frac{\mathcal{Z}(s)}{\text{6}\pi R}
\end{equation}
By adopting Eq. \eqref{eq:Eq40}, \citet{MasonWeitz1995} derived Eq. \eqref{eq:Eq37}. A rigorous derivation of Eq. \eqref{eq:Eq40} will require the solution of the appropriate three-dimensional continuum mechanics equations around the microsphere moving within the viscoelastic material. In the \citet{MasonWeitz1995} study, the mechanical behavior of the surrounding viscoelastic material was the unknown, so this challenge was bypassed by adopting a viscous--viscoelastic analogy where the Stokes' law when expressed in the Laplace space, $\mathcal{F}(s)=\text{6}\pi R \eta v(s)$ it can be generalized to $\mathcal{F}(s)=\text{6}\pi R \eta_{\text{VE}}(s) v(s)$ with Eq. \eqref{eq:Eq40}. This physically-motivated viscous--viscoelastic analogy adopted by \citet{MasonWeitz1995} is equivalent to the viscous--viscoelastic correspondence principle illustrated in Fig. \ref{fig:Fig04} which was conceived herein after observing the mathematical structure of Eqs. \eqref{eq:Eq03} and \eqref{eq:Eq15} in association with the concept of the inerter. 

The viscous--viscoelastic correspondence principle for Brownian motion simplifies appreciably the calculations of the mean squared displacement and of the velocity autocorrelation function of Brownian particles suspended in viscoelastic materials where inertia effects are non-negligible at longer time-scales as in the case of Brownian particles suspended in a Maxwell fluid.

\section{Brownian Motion within a Maxwell Fluid}
The Brownian motion of particles suspended in a Maxwell fluid with a single relaxation time, $\lambda=$ {\large $\frac{\eta}{G}$}, when subjected to the random forces, $f_R(t)$, is described by the Langevin equation \eqref{eq:Eq38}, where the relaxation kernel, $\zeta(t-\xi)$ is \citep{VolkovVinogradov1984, RodriguezSalinas1988, VolkovLeonov1996}
\begin{equation}\label{eq:Eq41}
\zeta(t-\xi)=\text{6}\pi R G_{\text{VE}}(t-\xi)=\text{6}\pi R G e^{-\frac{G}{\eta}(t-\xi)}
\end{equation}
where $G_{\text{VE}}(t-\xi)=G e${\large $^{-\frac{G}{\eta}(t-\xi)}$} is the relaxation modulus $($stress due to a unit-amplitude step--strain $\gamma(t)=U(t-\text{0}))$ of the Maxwell fluid \citep{BirdArmstrongHassager1987}. Equation \eqref{eq:Eq38} in association with Eq. \eqref{eq:Eq41} leads to the temporal evaluation of the particle's velocity autocorrelation functions \citep{VolkovLeonov1996} from which the mean squared displacement, $\left\langle \Delta r^{\text{2}}(t)  \right\rangle$, can be computed \citep{vanZantenRufener2000}.

\begin{figure}[b!]
\centering
\includegraphics[width=0.95\linewidth, angle=0]{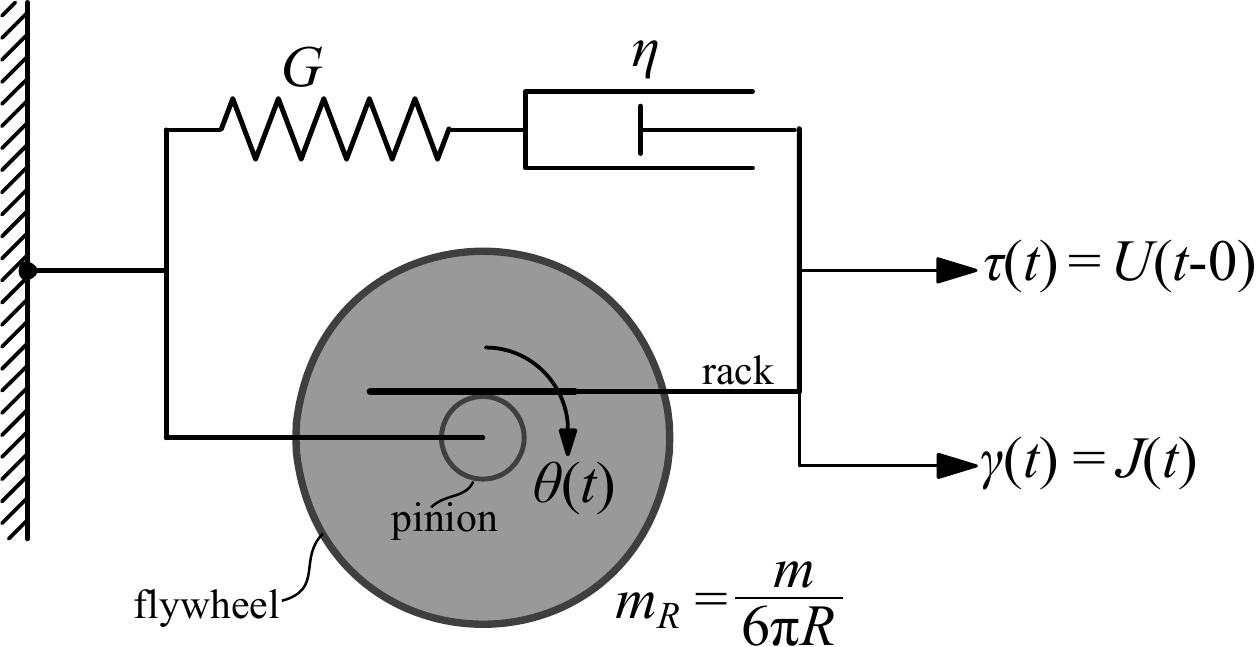}
\caption{Mechanical analogue for Brownian motion in a Maxwell fluid. It consists of the Maxwell fluid with shear modulus $G$ and viscosity $\eta$ that is connected in parallel with an inerter with distributed inertance $m_R=$ {\large $\frac{m}{\text{6}\pi R}$}.}
\label{fig:Fig05}
\end{figure}

In this Section the mean squared displacement of Brownian particles suspended in a Maxwell fluid is calculated with the correspondence principle summarized in Fig. \ref{fig:Fig04}. Accordingly, the problem reduces to the calculation of the creep compliance of a Maxwell fluid with shear modulus $G$ and viscosity $\eta$ that is connected in parallel with an inerter with distributed inertance $m_R=$ {\large $\frac{m}{\text{6}\pi R}$} as shown in Fig. \ref{fig:Fig05}.

The total stress $\tau(t)=\tau_{\text{1}}(t)+\tau_{\text{2}}(t)$ from the linear network shown in Fig. \ref{fig:Fig05} is the summation of the stress output from the Maxwell element, $\tau_{\text{1}}(t)$
\begin{equation}\label{eq:Eq42}
\tau_{\text{1}}(t)+\frac{\eta}{G} \frac{\mathrm{d}\tau_{\text{1}}(t)}{\mathrm{d}t}=\eta\frac{\mathrm{d}\gamma(t)}{\mathrm{d}t}
\end{equation}
and the stress output from the inerter, $\tau_{\text{2}}(t)$
\begin{equation}\label{eq:Eq43}
\tau_{\text{2}}(t)=m_R\frac{\mathrm{d}^{\text{2}}\gamma(t)}{\mathrm{d}t^{\text{2}}}
\end{equation}
The summation of Eqs. \eqref{eq:Eq42} and \eqref{eq:Eq43} together with the time-derivative of Eq. \eqref{eq:Eq43} yields a 3$^{\text{rd}}$-order constitutive equation for the linear network shown in Fig. \ref{fig:Fig05}
\begin{equation}\label{eq:Eq44}
\tau(t)+\frac{\eta}{G} \frac{\mathrm{d}\tau(t)}{\mathrm{d}t}=\eta\frac{\mathrm{d}\gamma(t)}{\mathrm{d}t} + m_R\frac{\mathrm{d}^{\text{2}}\gamma(t)}{\mathrm{d}t^{\text{2}}} + \frac{\eta \, m_R}{G}\frac{\mathrm{d}^{\text{3}}\gamma(t)}{\mathrm{d}t^{\text{3}}}
\end{equation}
By defining the dissipation time $\tau=$ {\large $\frac{m_R}{\eta}$} $=$ {\large $\frac{m}{\text{6}\pi R\eta}$} and the rotational angular frequency $\omega_R=$ {\large $\sqrt{\frac{G}{m_R}}$} $=$ {\large $\sqrt{\frac{\text{6}\pi R G}{m}}$}, Eq. \eqref{eq:Eq44} assumes the form
\begin{align}\label{eq:Eq45}
\tau(t)+ & \frac{\text{1}}{\tau \omega_R^{\text{2}}} \frac{\mathrm{d}\tau(t)}{\mathrm{d}t}= \\ \nonumber
& m_R \left( \frac{\text{1}}{\tau} \frac{\mathrm{d}\gamma(t)}{\mathrm{d}t} + \frac{\mathrm{d}^{\text{2}}\gamma(t)}{\mathrm{d}t^{\text{2}}} +\frac{\text{1}}{\tau \omega_R^{\text{2}}} \frac{\mathrm{d}^{\text{3}}\gamma(t)}{\mathrm{d}t^{\text{3}}} \right)
\end{align}
Laplace transform of Eq. \eqref{eq:Eq45} gives $\gamma(s)=\mathcal{J}(s)\tau(s)$, where $\mathcal{J}(s)$ is the complex dynamic compliance of the linear network shown in Fig. \ref{fig:Fig05}
\begin{equation}\label{eq:Eq46}
\mathcal{J}(s)=\frac{\text{1}}{\mathcal{G}(s)}=\frac{\gamma(s)}{\tau(s)}=\frac{\text{1}}{m_R} \, \frac{\text{1}+\frac{\text{1}}{\tau\omega_R^{\text{2}}}s}{s\left( \frac{\text{1}}{\tau} + s + \frac{\text{1}}{\tau\omega_R^{\text{2}}}s^{\text{2}} \right)}
\end{equation}
In addition to $s=$ 0, the other two poles of the complex dynamic compliance, $\mathcal{J}(s)$ given by Eq. \eqref{eq:Eq46} are
\begin{equation}\label{eq:Eq47}
s_{\text{1}}=-\frac{\tau\omega_R^{\text{2}}}{\text{2}}+\omega_R\sqrt{\left( \frac{\tau\omega_R}{\text{2}} \right)^{\text{2}}-\text{1}}=-\omega_R\left( \beta-\sqrt{\beta^{\text{2}}-\text{1}} \right)
\end{equation}
and
\begin{equation}\label{eq:Eq48}
s_{\text{2}}=-\frac{\tau\omega_R^{\text{2}}}{\text{2}}-\omega_R\sqrt{\left( \frac{\tau\omega_R}{\text{2}} \right)^{\text{2}}-\text{1}}=-\omega_R\left( \beta+\sqrt{\beta^{\text{2}}-\text{1}} \right)
\end{equation}
where $\beta=$ {\large $\frac{\tau\omega_R}{\text{2}}$} $=$ {\large $\frac{\text{1}}{\text{2}\eta}\sqrt{\frac{mG}{\text{6}\pi R}}$}  is a dimensionless parameter of the linear network and of the Brownian particle--Maxwell fluid system. By virtue of Eqs. \eqref{eq:Eq47} and \eqref{eq:Eq48}, the complex creep function $\mathcal{C}(s)=$ {\large $\frac{\mathcal{J}(s)}{s}$} $=$ $\displaystyle\int_{\text{0}}^{\infty}J(t)e^{-st}\mathrm{d}t$ is expressed as
{\small %
\begin{align}\label{eq:Eq49}
\mathcal{C}(s)=\frac{\mathcal{J}(s)}{s}=\frac{\text{1}}{\eta} \Bigg[ \frac{\text{1}}{s^{\text{2}}} & - \left( \frac{\text{2}\beta^{\text{2}}-\text{1}}{\text{4}\beta\sqrt{\beta^{\text{2}}-\text{1}}} + \frac{\text{1}}{\text{2}}\right)\frac{\text{1}}{s(s-s_{\text{1}})} + \\ \nonumber 
& \left( \frac{\text{2}\beta^{\text{2}}-\text{1}}{\text{4}\beta\sqrt{\beta^{\text{2}}-\text{1}}} -\frac{\text{1}}{\text{2}}\right)\frac{\text{1}}{s(s-s_{\text{2}})} \Bigg]
\end{align}}
For the case where $\beta=$ {\large $\frac{\tau\omega_R}{\text{2}}$} $>$ 1 (stiff spring), inverse Laplace transform of Eq. \eqref{eq:Eq49} gives
{\small %
\begin{align}\label{eq:Eq50}
J(t)=& \mathcal{L}^{-\text{1}}\left\lbrace \mathcal{C}(s) \right\rbrace =  \frac{m_R}{\eta^{\text{2}}} \Bigg\lbrace \frac{t}{\tau} - \frac{\text{1}}{\text{4}\beta^{\text{2}}} \Bigg[ \text{4}\beta^{\text{2}} -\text{1} - \\ \nonumber
& e^{-\text{2}\frac{t}{\tau}\beta^{\text{2}}} \Bigg( \frac{\text{4}\beta^{\text{3}}-\text{3}\beta}{\sqrt{\beta^{\text{2}}-\text{1}}}\sinh\left( \text{2}\frac{t}{\tau}\beta \sqrt{\beta^{\text{2}}-\text{1}} \right) + \\ 
& \left( \text{4}\beta^{\text{2}}-\text{1} \right) \cosh\left( \text{2}\frac{t}{\tau}\beta \sqrt{\beta^{\text{2}}-\text{1}}  \right)  \Bigg) \Bigg] \Bigg\rbrace \text{, } \enskip  \beta > \text{1;} \nonumber
\end{align}}
whereas, for the case where $\beta=$ {\large $\frac{\tau\omega_R}{\text{2}}$} $<$ 1 (flexible spring), the creep compliance of the network shown in Fig. \ref{fig:Fig05} is
{\small %
\begin{align}\label{eq:Eq51}
J(t)=& \mathcal{L}^{-\text{1}}\left\lbrace \mathcal{C}(s) \right\rbrace =  \frac{m_R}{\eta^{\text{2}}} \Bigg\lbrace \frac{t}{\tau} - \frac{\text{1}}{\text{4}\beta^{\text{2}}} \Bigg[ \text{4}\beta^{\text{2}} -\text{1} - \\ \nonumber
& e^{-\text{2}\frac{t}{\tau}\beta^{\text{2}}} \Bigg( \frac{\text{4}\beta^{\text{3}}-\text{3}\beta}{\sqrt{\text{1}-\beta^{\text{2}}}}\sin\left( \text{2}\frac{t}{\tau}\beta \sqrt{\text{1}-\beta^{\text{2}}} \right) + \\ \nonumber
& \left( \text{4}\beta^{\text{2}}-\text{1} \right) \cos\left( \text{2}\frac{t}{\tau}\beta \sqrt{\text{1}-\beta^{\text{2}}}  \right)  \Bigg) \Bigg] \Bigg\rbrace \text{, } \enskip  \beta < \text{1} \nonumber
\end{align}}
By employing the correspondence principle for Brownian motion, the mean squared displacement of Brownian particles suspended in a Maxwell fluid with elasticity $G$ and viscosity $\eta$ is given by Eq. \eqref{eq:Eq12} where now $J(t)$ is offered by Eqns. \eqref{eq:Eq50} or \eqref{eq:Eq51}. Figure \ref{fig:Fig06} plots the normalized mean squared displacement for Brownian motion in a Maxwell fluid
\begin{equation}\label{eq:Eq52}
\frac{\text{18}\pi^{\text{2}}R^{\text{2}}\eta^{\text{2}}}{m N K_B T} \left\langle \Delta r^{\text{2}}(t) \right\rangle = \frac{\eta^{\text{2}}}{m_R}J(t)
\end{equation}
\begin{figure}[b!]
\centering
\includegraphics[width=\linewidth, angle=0]{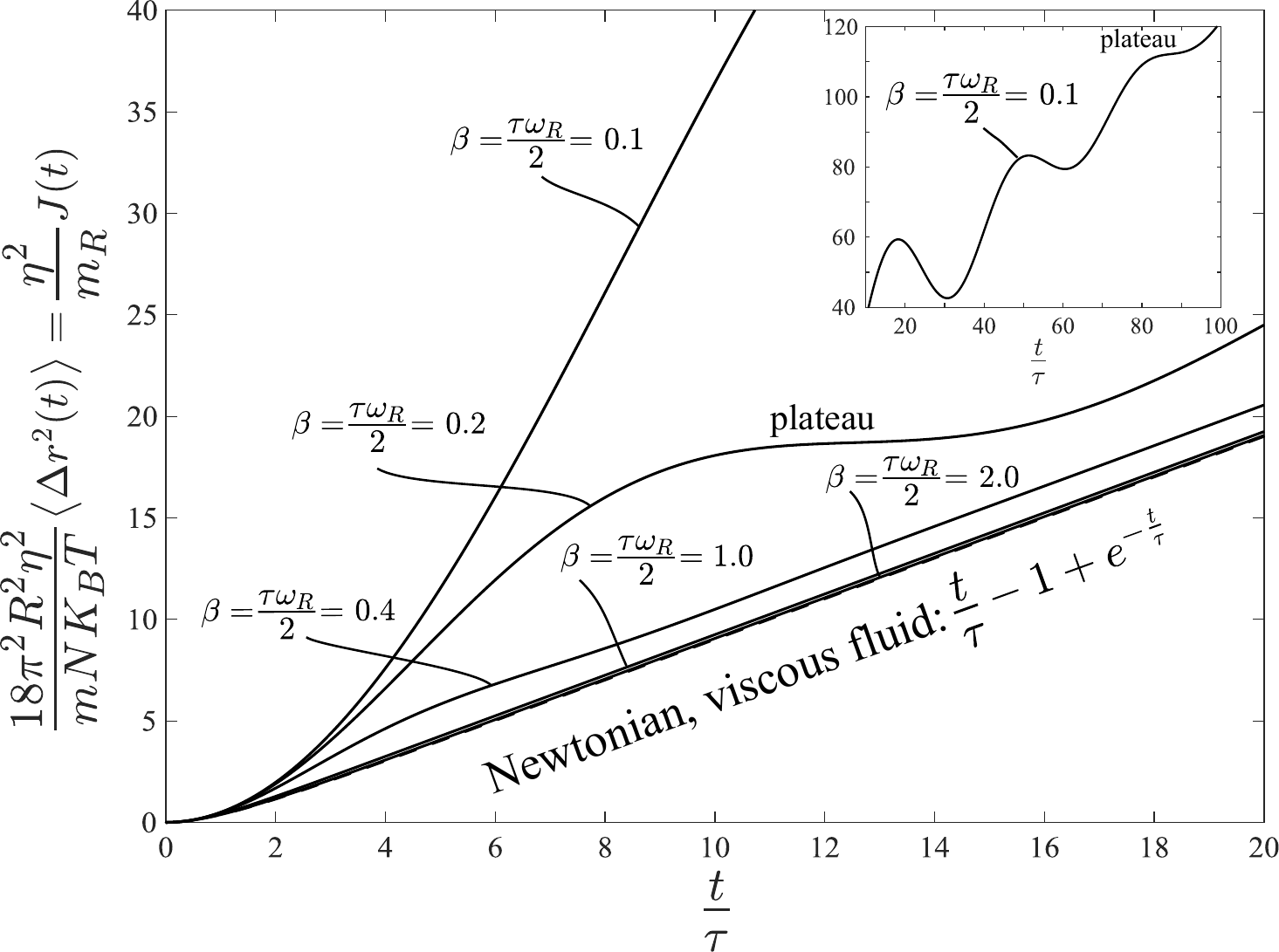}
\caption{Normalized mean squared displacement of Brownian particles suspended in a Maxwell fluid with shear modulus $G$ and viscosity $\eta$ for various values of the parameter $\beta=$ {\large $\frac{\text{1}}{\text{2}\eta}\sqrt{\frac{mG}{\text{6}\pi R}}$} $=$ {\large $\frac{\text{1}}{\text{2}}\frac{\sqrt{m_R}\sqrt{G}}{\eta}$}. For values of $\beta<$ 1 (weak spring), the inertia effects are more pronounced and the mean squared displacement curves exhibit a reversal of curvature as time increases together with a plateau $\Big( \beta=$ 0.2 at {\large $\frac{t}{\tau}$} $\approx$ 12 and $\beta=$ 0.1 at {\large $\frac{t}{\tau}$} $\approx$ 90 shown in the inset plot$\Big)$.}
\label{fig:Fig06}
\end{figure}
as a function of the dimensionless time {\large $\frac{t}{\tau}$} for various values of $\beta=$ {\large $\frac{\tau\omega_R}{\text{2}}$} $=$ {\large $\frac{\text{1}}{\text{2}\eta}\sqrt{\frac{mG}{\text{6}\pi R}}$} $=$ {\large $\frac{\text{1}}{\text{2}}\frac{\sqrt{m_R}\sqrt{G}}{\eta}$}. 

For values of $\beta <$ 1, the shear modulus $G$ is weak; therefore, the inertia effects are more pronounced. In this case the mean squared displacement shown in Fig. \ref{fig:Fig06} exhibits a reversal of curvature as the dimensionless time {\large $\frac{t}{\tau}$} increases; while for selective values of $\beta$ (say $\beta=$ 0.1, 0.2) it exhibits a plateau \citep{vanZantenRufener2000}. This distinct behavior at time-regimes which are orders of magnitude larger than $\tau=$ {\large $\frac{m}{\text{6}\pi R \eta}$}, is due to the inertia effects and cannot be captured by the approximate Eq. \eqref{eq:Eq02} proposed in \cite{PalmerXuWirtz1998,XuViasnoffWirtz1998} since the creep compliance of the surrounding viscoelastic Maxwell fluid is merely a linear function of time, $J_{\text{VE}}(t)=$ {\large $\frac{\text{1}}{G}$} $+$ {\large $\frac{\text{1}}{\eta}$}$t$ with $t\geq \text{0}$. As the value of  $\beta=$ {\large $\frac{\tau\omega_R}{\text{2}}$} increases $($stiffer $G)$, the inertia effects are suppressed and the normalized mean squared displacement of Brownian particles suspended in a Maxwell fluid tends to that of Brownian particles suspended in a Newtonian, viscous fluid: {\large $\frac{t}{\tau}$}$-\text{1}+e${\large $^{-\frac{t}{\tau}}$}.

The velocity autocorrelation function of Brownian particles suspended in a Maxwell fluid is offered by Eq. \eqref{eq:Eq34}, where $\psi(t) = \mathcal{L}^{-\text{1}}\left\lbrace \phi(s) \right\rbrace$ is the impulse strain-rate response function of the mechanical network shown in Fig. \ref{fig:Fig05}. The complex dynamic fluidity $\phi(s)=$ {\large $\frac{s}{\mathcal{G}(s)}$} of the mechanical network shown in Fig. \ref{fig:Fig05} derives directly from Eq. \eqref{eq:Eq46}
\begin{equation}\label{eq:Eq53}
\phi(s)=\frac{\text{1}}{m_R}\,\frac{s+\tau\omega_R^{\text{2}}}{(s-s_{\text{1}})(s-s_{\text{2}})}
\end{equation}
where the poles $s_{\text{1}}$ and $s_{\text{2}}$ are given by Eqs. \eqref{eq:Eq47} and \eqref{eq:Eq48}. For the case where $\beta=$ {\large $\frac{\tau\omega_R}{\text{2}}$} $>$ 1 (stiff spring), inverse Laplace transform of Eq. \eqref{eq:Eq53} gives
{\small %
\begin{align}\label{eq:Eq54}
\psi(t)= & \mathcal{L}^{-\text{1}}\left\lbrace \phi(s) \right\rbrace= \frac{\text{1}}{m_R}  e^{-\text{2} \frac{t}{\tau}\beta^{\text{2}}} \Bigg[  \cosh\left( \text{2}\frac{t}{\tau}\beta\sqrt{\beta^{\text{2}}-\text{1}} \right) + \\ \nonumber
&\frac{\beta}{\sqrt{\beta^{\text{2}}-\text{1}}} \sinh\left( \text{2}\frac{t}{\tau}\beta\sqrt{\beta^{\text{2}}-\text{1}} \right) \Bigg]  \text{, } \enskip  \beta > \text{1;}
\end{align}}
whereas, for the case where $\beta=$ {\large $\frac{\tau\omega_R}{\text{2}}$} $<$ 1 (flexible spring), the impulse strain-rate response function of the mechanical network shown in Fig. \ref{fig:Fig05} is
{\small %
\begin{align}\label{eq:Eq55}
\psi(t)= & \mathcal{L}^{-\text{1}}\left\lbrace \phi(s) \right\rbrace=\frac{\text{1}}{m_R}  e^{-\text{2} \frac{t}{\tau}\beta^{\text{2}}} \Bigg[ \cos\left( \text{2}\frac{t}{\tau}\beta\sqrt{\text{1}-\beta^{\text{2}}} \right) + \\ \nonumber
&\frac{\beta}{\sqrt{\text{1}-\beta^{\text{2}}}} \sin\left( \text{2}\frac{t}{\tau}\beta\sqrt{\text{1}-\beta^{\text{2}}} \right) \Bigg]  \text{, } \enskip  \beta < \text{1}
\end{align}}

By employing the correspondence principle for Brownian motion, the velocity autocorrelation function of Brownian particles suspended in a Maxwell fluid with elasticity $G$ and viscosity $\eta$ is given by Eq. \eqref{eq:Eq34} where now $\psi(t)$ is offered by Eqs. \eqref{eq:Eq54} or \eqref{eq:Eq55}. Figure \ref{fig:Fig07} plots the normalized velocity autocorrelation function for Brownian motion in a Maxwell fluid
\begin{figure}[b!]
\centering
\includegraphics[width=\linewidth, angle=0]{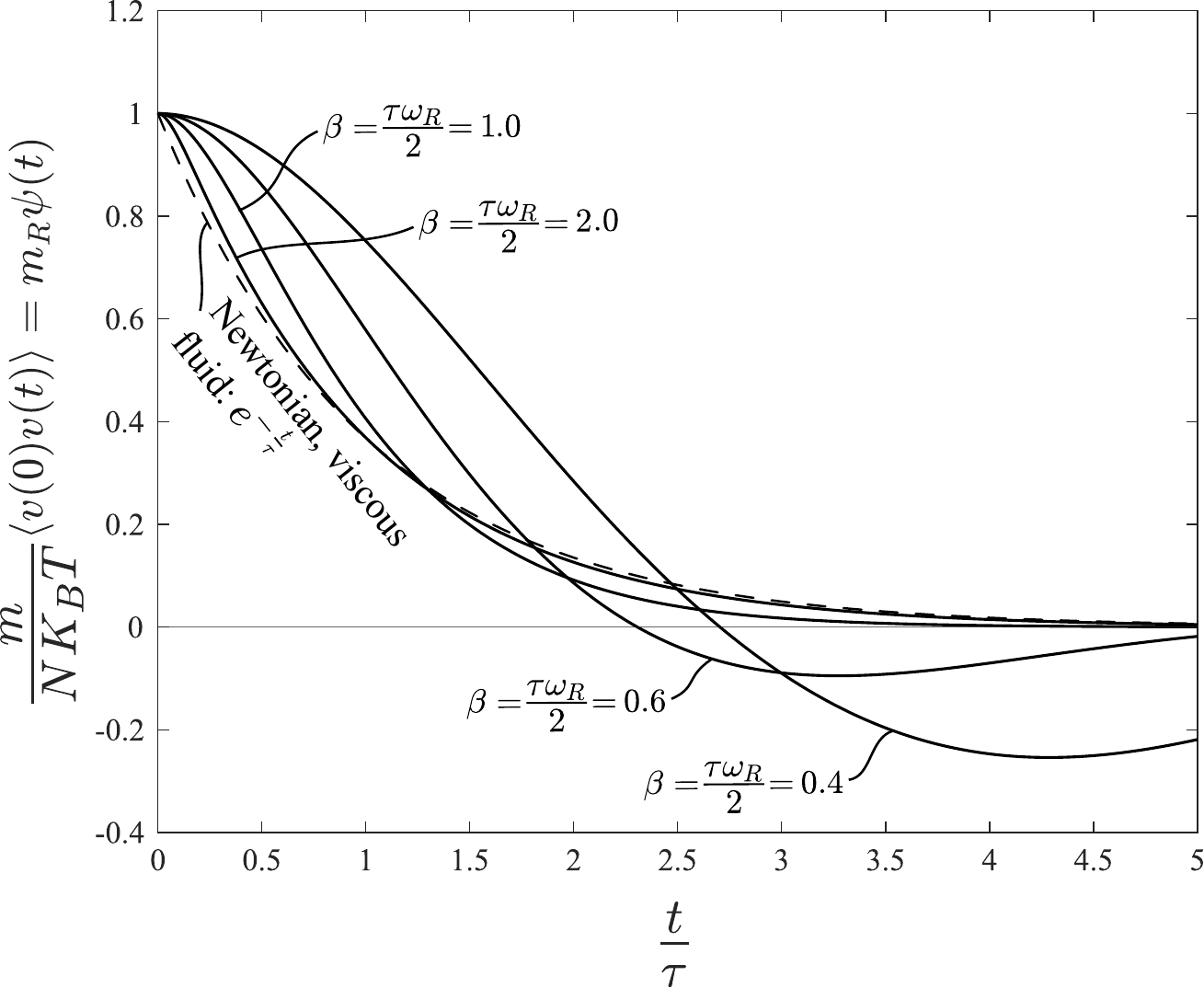}
\caption{Normalized velocity autocorrelation function of Brownian particles suspended in a Maxwell fluid with shear modulus $G$ and viscosity $\eta$ for various values of the parameter $\beta=$ {\large $\frac{\text{1}}{\text{2}\eta}\sqrt{\frac{mG}{\text{6}\pi R}}$} $=$ {\large $\frac{\text{1}}{\text{2}}\frac{\sqrt{m_R}\sqrt{G}}{\eta}$}. For values of $\beta>$ 1 (stiff spring), the velocity autocorrelation curves tend to the velocity autocorrelation function of Brownian particles in a Newtonian viscous fluid: $e${\large $^{-\frac{t}{\tau}}$}.}
\label{fig:Fig07}
\end{figure}
\begin{equation}\label{eq:Eq56}
\frac{m}{N K_B T}\left\langle v(\text{0})v(t) \right\rangle = m_R \psi(t)
\end{equation}
As the value of $\beta=$ {\large $\frac{\tau\omega_R}{\text{2}}$} increases (stiffer $G$), the normalized velocity autocorrelation function of Brownian particles suspended in a Maxwell fluid tends to that of Brownian particles suspended in a Newtonian viscous fluid: $e$ {\large $^{-\frac{t}{\tau}}$}.

\section{Brownian Motion within a Subdiffusive Material}
Most materials exhibit both viscous and elastic behavior. For such materials the thermally driven motion of embedded microspheres reflects the combined viscous and elastic contributions which are revealed in the time evolution of the mean squared displacement and the velocity autocorrelation function as shown in Figs. \ref{fig:Fig06} and \ref{fig:Fig07} for the case where the probe microspheres are suspended in a Maxwell fluid. Several complex materials exhibit a subdiffusive behavior where from early times and over several temporal decades the mean squared displacement grows with time according to a power law, $\left\langle \Delta r^{\text{2}}(t)  \right\rangle \sim t^{\alpha}$, where 0 $\leq \alpha \leq$ 1 is the diffusive exponent \citep{PalmerXuWirtz1998, XuViasnoffWirtz1998, GislerWeitz1999}.

Early studies on the behavior of viscoelastic materials that their time--response functions follow power laws have been presented by \citet{Nutting1921}, who noticed that the stress response of several fluid-like materials to a step strain decays following a power law, $\tau(t)=G_{\text{VE}}(t)\sim t^{- \alpha}$ with 0 $\leq \alpha \leq$ 1. Following \citeauthor{Nutting1921}'s observations and the early work of \citet{Gemant1936, Gemant1938} on fractional differentials, \citet{ScottBlair1944, ScottBlair1947} pioneered the introduction of fractional calculus in viscoelasticity. With analogy to the Hookean spring, in which the stress is proportional to the zero-th derivative of the strain and the Newtonian dashpot, in which the stress is proportional to the first derivative of the strain, Scott Blair and his co-workers \cite{ScottBlair1944, ScottBlair1947, ScottBlairCaffyn1949} proposed the springpot element --- that is a mechanical element in-between a spring and a dashpot with constitutive law
\begin{equation}\label{eq:Eq57}
\tau(t)=\mu_{\alpha} \frac{\mathrm{d}^{\alpha}\gamma(t)}{\mathrm{d}t^{\alpha}} \text{,} \enskip \text{0} \leq \alpha \leq \text{1} 
\end{equation}
where $\alpha$ is a positive real number, 0 $\leq \alpha \leq$ 1, $\mu_{\alpha}$ is a phenomenological material parameter with units $\left[\text{M}\right]\left[\text{L}\right]^{-\text{1}}\left[\text{T}\right]^{\alpha-\text{2}}$ (say \textit{Pa$\cdot$sec}$^{\alpha}$) and {\large $\frac{\mathrm{d}^{\alpha}\gamma(t)}{\mathrm{d}t^{\alpha}}$} is the fractional derivative of order $\alpha$ of the strain history, $\gamma(t)$.

A definition of the fractional derivative of order $\alpha$ is given through the convolution integral
\begin{equation}\label{eq:Eq58}
I^{\alpha}\gamma(t)=\frac{\text{1}}{\Gamma(\alpha)}\int_{c}^{t}(t-\xi)^{\alpha-\text{1}}\gamma(\xi)\mathrm{d}\xi
\end{equation}
where $\Gamma(\alpha)$ is the Gamma function. When the lower limit, $c=\text{0}$, the integral given by Eq. \eqref{eq:Eq58} is often referred to as the Riemann-Liouville fractional integral \citep{OldhamSpanier1974, SamkoKilbasMarichev1974, MillerRoss1993, Podlubny1998}. The integral in Eq. \eqref{eq:Eq58} converges only for $\alpha>\text{0}$, or in the case where $\alpha$ is a complex number, the integral converges for $\mathcal{R}(\alpha)>\text{0}$. Nevertheless, by a proper analytic continuation across the line $\mathcal{R}(\alpha)=\text{0}$, and provided that the function $\gamma(t)$ is $n$ times differentiable, it can be shown that the integral given by Eq. \eqref{eq:Eq58} exists for $n-\mathbb{R}(\alpha)>\text{0}$ \citep{Riesz1949}. In this case the fractional derivative of order $\alpha\in \mathbb{R}^+$ exists and is defined as
\begin{equation}\label{eq:Eq59}
\frac{\mathrm{d}^{\alpha}\gamma(t)}{\mathrm{d}t^{\alpha}}=I^{-\alpha}\gamma(t)=\frac{\text{1}}{\Gamma(-\alpha)}\int_{\text{0}^-}^{t}\frac{\gamma(\xi)}{(t-\xi)^{\alpha+\text{1}}} \mathrm{d}\xi  \text{,} \enskip \alpha\in \mathbb{R}^+
\end{equation}
where $\mathbb{R}^+$ is the set of positive real numbers and the lower limit of integration, $\text{0}^-$, may engage an entire singular function at the time origin such as $\gamma(t)=\delta(t-\text{0})$ \citep{Lighthill1958}. Eq. \eqref{eq:Eq59} indicates that the fractional derivative of order $\alpha$ of $\gamma(t)$ is essentially the convolution of $\gamma(t)$ with the kernel {\large $\frac{t^{-\alpha-\text{1}}}{\Gamma(-\alpha)}$} \citep{OldhamSpanier1974, SamkoKilbasMarichev1974, MillerRoss1993, Mainardi2010}. The Riemann--Liouville definition of the fractional derivative of order $\alpha\in \mathbb{R}^+$ given by Eq. \eqref{eq:Eq59}, where the lower limit of integration is zero, is relevant to rheology since the strain and stress histories, $\gamma(t)$ and $\tau(t)$, are causal functions, being zero at negative times.  

The relaxation modulus $($stress history due to a unit-amplitude step--strain, $\gamma(t)=U(t-\text{0}))$ of the springpot element (Scott-Blair fluid) expressed by Eq. \eqref{eq:Eq57} is \citep{SmitDeVries1970, Koeller1984, Friedrich1991, HeymansBauwens1994, SchiesselMetzlerBlumenNonnenmacher1995, PaladeVerneyAttane1996}
\begin{equation}\label{eq:Eq60}
G_{\text{VE}}(t)=\mu_{\alpha}\frac{\text{1}}{\Gamma(\text{1}-\alpha)}t^{-\alpha} \text{,} \enskip t> \text{0}
\end{equation}
which decays by following the power-law initially observed by \citet{Nutting1921}. The creep compliance (retardation function) of the springpot element is \citep{Koeller1984, Friedrich1991, HeymansBauwens1994, SchiesselMetzlerBlumenNonnenmacher1995}
\begin{equation}\label{eq:Eq61}
J_{\text{VE}}(t)=\frac{\text{1}}{\mu_{\alpha}}\frac{\text{1}}{\Gamma(\text{1}+\alpha)}t^{\alpha} \text{,} \enskip t\geq \text{0}
\end{equation}
The power-law, $t^{\alpha}$, appearing in Eq. \eqref{eq:Eq61} renders the elementary springpot element expressed by Eq. \eqref{eq:Eq57} (Scott-Blair fluid) a suitable phenomenological model to study Brownian motion in subdiffusive materials.

The Brownian motion of a microsphere suspended in a subdiffusive material that its viscoelastic behavior is approximated with the Scott-Blair (springpot) element with relaxation modulus $G_{\text{VE}}(t)$ given by Eq. \eqref{eq:Eq60} is described by the Langevin equation \eqref{eq:Eq38} where the relaxation kernel $\zeta(t-\xi)$ is \citep{Lutz2001, DespositoVinales2009}
\begin{equation}\label{eq:Eq62}
\zeta(t-\xi)=\text{6}\pi R G_{\text{VE}}(t-\xi)=\text{6}\pi R \mu_{\alpha} \frac{\text{1}}{\Gamma(\text{1}-\alpha)}\frac{\text{1}}{(t-\xi)^{\alpha}}
\end{equation}
and the Langevin equation assumes the expression
\begin{equation}\label{eq:Eq63}
m\frac{\mathrm{d}v(t)}{\mathrm{d}t}+\text{6}\pi R \mu_{\alpha} \int_{\text{0}^-}^{t} \frac{\text{1}}{\Gamma(\text{1}-\alpha)}\frac{\text{1}}{(t-\xi)^{\alpha}}v(\xi)\mathrm{d}\xi=f_R(t)
\end{equation}
When the order of differentiation in Eq. \eqref{eq:Eq57} is $\alpha=$ 1, the springpot element becomes a Newtonian dashpot with $\mu_\alpha=\mu_{\text{1}}=\eta$; while the kernel {\large $\frac{\text{1}}{\Gamma(\text{1}-\alpha)}\frac{\text{1}}{(t-\xi)^{\alpha}}$} becomes the Dirac delta function, $\delta(t-\text{0})$ according to the \citet{GelfandShilov1964} definition of the Dirac delta function and its integer-order derivatives
\begin{equation}\label{eq:Eq64}
\frac{\mathrm{d}^n\delta(t-\xi)}{\mathrm{d}t^n}=\frac{\text{1}}{\Gamma(-n)}\frac{\text{1}}{(t-\xi)^{n+\text{1}}} \text{,} \enskip \text{with} \enskip n \in \mathbb{N}_{\text{0}} 
\end{equation}
where $\mathbb{N}_{\text{0}}$ is the set of positive integers including zero. Accordingly, for the limiting case where $\alpha=$ 1, the convolution in the Langevin equation \eqref{eq:Eq63} reduces to the Stokes term
\begin{equation}\label{eq:Eq65}
\text{6}\pi R\eta\int_{\text{0}^-}^{t} \delta(t-\xi)v(\xi)\mathrm{d}\xi=\text{6}\pi R\eta v(t)
\end{equation}
and Eq. \eqref{eq:Eq63} contracts to Eq. \eqref{eq:Eq04}. At the other limit where $\alpha=$ 0 in Eq. \eqref{eq:Eq57}, the springpot element becomes a Hookean spring with elastic shear modulus $\mu_\alpha=\mu_{\text{0}}=G$, while the kernel {\large $\frac{\text{1}}{\Gamma(\text{1})}\frac{\text{1}}{(t-\xi)^{\text{0}}}$} $=$ 1. Accordingly, for the limiting case where $\alpha=$ 0, the convolution in the Langevin equation \eqref{eq:Eq63} reduces to
\begin{equation}\label{eq:Eq66}
\text{6}\pi R G \int_{\text{0}}^{t}  v(\xi) \mathrm{d}\xi = \text{6}\pi R G r(t)
\end{equation}
where $r(t)$ is the displacement of the microsphere and $\text{6}\pi R G=k$ is the spring constant of the restoring force on the microsphere \citep{SchnurrGittesMacKintoshSchmidt1997}. In this limiting case $(\alpha=\text{0})$, Eq. \eqref{eq:Eq63} describes the Brownian motion of the particle in a harmonic trap with zero damping (see Eq. \eqref{eq:Eq13} with $\zeta=$ 0).

With reference to Eq. \eqref{eq:Eq58}, the convolution integral appearing in Eq. \eqref{eq:Eq63} is the fractional integral of order 1 $-\alpha$ of the velocity history; therefore, the Langevin equation \eqref{eq:Eq63} can be expressed in a compact form
\begin{equation}\label{eq:Eq67}
m\frac{\mathrm{d}v(t)}{\mathrm{d}t}+ \text{6}\pi R \mu_{\alpha} I^{\text{1}-\alpha}v(t)=f_R(t)
\end{equation}
which is the anticipated result given the fractional derivative constitutive equation \eqref{eq:Eq57} of the surrounding viscoelastic material.

The mean squared displacement of a Brownian particle suspended in the fractional Scott-Blair fluid described by Eq. \eqref{eq:Eq57} has been evaluated in \cite{KobelevRomanov2000, Lutz2001} after computing the velocity autocorrelation function of the random process described by Eq. \eqref{eq:Eq63}
\begin{equation}\label{eq:Eq68}
\left\langle \Delta r^{\text{2}}(t) \right\rangle = \frac{\text{2}N K_B T}{m}t^{\text{2}}E_{\text{2}-\alpha\text{, 3}}\left( - \frac{\text{6}\pi R \mu_{\alpha}}{m} t^{\text{2}-\alpha}\right)
\end{equation}
where $E_{\alpha\text{, }\beta}(z)$ is the two-parameter Mittag--Leffler function \citep{Erdelyi1953, GorenfloKilbasMainardiRogosin2014}
\begin{equation}\label{eq:Eq69}
E_{\alpha\text{, }\beta}(z)=\sum\limits_{j=\text{0}}^{\infty}\frac{z^j}{\Gamma(j\alpha+\beta)} \text{,} \enskip \alpha \text{, } \beta > \text{0}
\end{equation}
Herein, the mean squared displacement of Brownian particles suspended in the fractional Scott-Blair fluid described by Eq. \eqref{eq:Eq57} is calculated with the correspondence principles summarized in Fig. \ref{fig:Fig04}. Accordingly, the problem reduces to the calculation of the creep compliance of the springpot element described by Eq. \eqref{eq:Eq57} that is connected in parallel with an inerter with distributed inertance $m_R=$ {\large $\frac{m}{\text{6}\pi R}$} as shown in Fig. \ref{fig:Fig01} in which the dashpot is replaced with a springpot. Given the parallel connection of the springpot and the inerter, the constitutive law is
\begin{equation}\label{eq:Eq70}
\tau(t)=\mu_{\alpha}\frac{\mathrm{d}^{\alpha}\gamma(t)}{\mathrm{d}t^{\alpha}}+ m_R\frac{\mathrm{d}^{\text{2}}\gamma(t)}{\mathrm{d}t^{\text{2}}} \text{,} \enskip \alpha \in \mathbb{R}^+
\end{equation}
The Laplace transform of Eq. \eqref{eq:Eq70} gives $\gamma(s)=\mathcal{J}(s)\tau(s)$, where $\mathcal{J}(s)$ is the complex dynamic compliance of the linear network
\begin{equation}\label{eq:Eq71}
\mathcal{J}(s)=\frac{\text{1}}{\mathcal{G}(s)}=\frac{\text{1}}{\mu_{\alpha}s^{\alpha}+m_Rs^{\text{2}}}=\frac{\text{1}}{m_R}\frac{\text{1}}{s^{\alpha}\left( s^{\text{2}-\alpha} + \frac{\mu_{\alpha}}{m_R} \right)}
\end{equation}
therefore, the complex creep function, $\mathcal{C}(s)=\displaystyle\int_{\text{0}}^{\infty}J(t)e^{-st}\mathrm{d}t$ of the springpot--inerter parallel connection is 
\begin{equation}\label{eq:Eq72}
\mathcal{C}(s)=\frac{\mathcal{J}(s)}{s}=\frac{\text{1}}{m_R} \frac{\text{1}}{s^{\alpha+\text{1}}\left( s^{\text{2}-\alpha} + \frac{\mu_{\alpha}}{m_R} \right)}
\end{equation}
The inverse Laplace transform of Eq. \eqref{eq:Eq72} is evaluated with the convolution integral \citep{Erdelyi1954}
\begin{equation}\label{eq:Eq73}
J(t)=\mathcal{L}^{-\text{1}}\left\lbrace \mathcal{C}(s) \right\rbrace = \int_{\text{0}}^{t}f(t-\xi)h(\xi)\mathrm{d}\xi
\end{equation}
with
\begin{equation}\label{eq:Eq74}
f(t)=\mathcal{L}^{-\text{1}}\left\lbrace \frac{\text{1}}{m_R} \frac{\text{1}}{s^{\alpha+\text{1}}} \right\rbrace = \frac{\text{1}}{m_R} \frac{\text{1}}{\Gamma(\text{1}+\alpha)}t^{\alpha}
\end{equation}
and 
\begin{equation}\label{eq:Eq75}
h(t)=\mathcal{L}^{-\text{1}}\left\lbrace \frac{\text{1}}{s^{\text{2}-\alpha}+\frac{\mu_{\alpha}}{m_R}} \right\rbrace = t^{\text{1}-\alpha} E_{\text{2}-\alpha\text{, 2}-\alpha}\left( - \frac{\mu_{\alpha}}{m_R} t^{\text{2}-\alpha} \right)
\end{equation}
\begin{figure}[b!]
\centering
\includegraphics[width=\linewidth, angle=0]{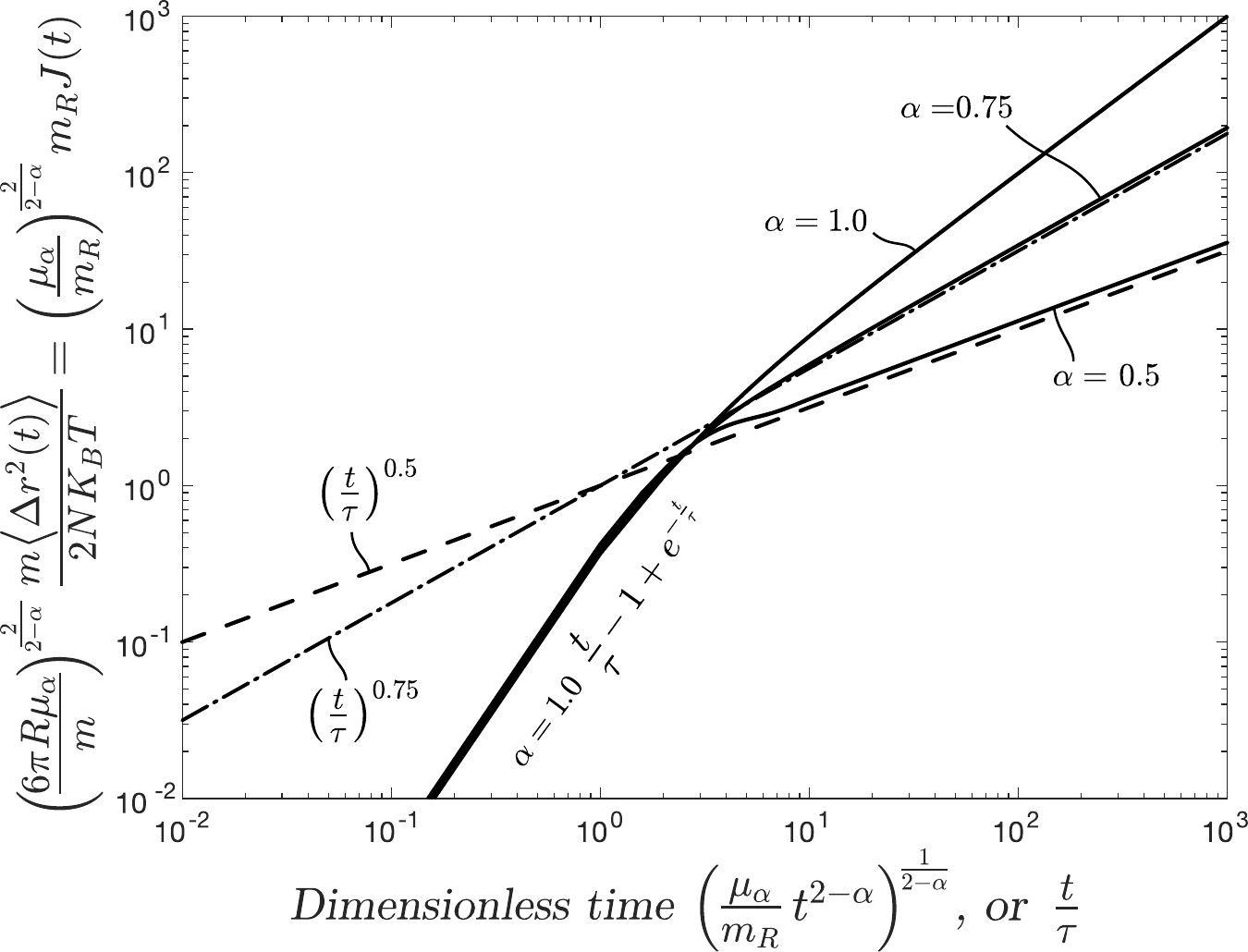}
\caption{Normalized mean squared displacement of Brownian particles suspended in a fractional subdiffusive Scott-Blair fluid with material constant $\mu_{\alpha}$ with units $\left[\text{M}\right]\left[\text{L}\right]^{-\text{1}}\left[\text{T}\right]^{\alpha-\text{2}}$ for various values of the fractional exponent, 0 $\leq \alpha \leq$ 1, as a function of the dimensionless time  $\Big(${\large $\frac{\mu_{\alpha}}{m_R}$}$t^{\text{2}-\alpha} \Big)${\large $^{\frac{\text{1}}{\text{2}-\alpha}}$} where $m_R=$ {\large $\frac{m}{\text{6}\pi R}$} and $\tau=$ {\large $\frac{m}{\text{6}\pi R \eta}$}.}
\label{fig:Fig08}
\end{figure}
where $E_{\alpha\text{, }\beta}(z)$ is the two-parameter Mittag--Leffler function defined by Eq. \eqref{eq:Eq69}. The function $h(t)$ expressed by Eq. \eqref{eq:Eq75} is also known in rheology as the Rabotnov function {\large $\varepsilon$}$_{\text{1}-\alpha}(-\lambda\text{, } t)=t^{\text{1}-\alpha}E_{\text{2}-\alpha\text{, 2}-\alpha}(-\lambda t^{\text{2}-\alpha})$ \citep{Mainardi2010, Rabotnov1980}. Substitution of the results of Eqs. \eqref{eq:Eq74} and \eqref{eq:Eq75} into the convolution given by Eq. \eqref{eq:Eq73}, the creep compliance of the springpot--inerter parallel connection is merely the fractional integral of order $\text{1} + \alpha$ of the Rabotnov function given by Eq. \eqref{eq:Eq75} \citep{MakrisEfthymiou2020}
\begin{align}\label{eq:Eq76}
J(t)= & \frac{\text{1}}{m_R} \frac{\text{1}}{\Gamma(\text{1}+\alpha)} \times \\ \nonumber 
& \int_{\text{0}}^{t}(t-\xi)^{\alpha}\xi^{\text{1}-\alpha}E_{\text{2}-\alpha\text{, 2}-\alpha} \left( -\frac{\mu_{\alpha}}{m_R} \xi^{\text{2}-\alpha} \right) \mathrm{d}\xi = \\ \nonumber
& \frac{\text{1}}{m_R} t^\text{2} E_{\text{2}-\alpha\text{, 3}}\left( - \frac{\mu_{\alpha}}{m_R} t^{\text{2}-\alpha} \right)
\end{align} 
By comparing the results of Eqs. \eqref{eq:Eq68} and \eqref{eq:Eq76} after using $m_R=$ {\large $\frac{m}{\text{6}\pi R}$}, the mean squared displacement $\left\langle \Delta r^{\text{2}}(t) \right\rangle$ of Brownian particles suspended in a subdiffusive fractional fluid is given again by Eq. \eqref{eq:Eq12} where now $J(t)$ is offered by Eq. \eqref{eq:Eq76}. Figure \ref{fig:Fig08} plots the normalized mean squared displacement for Brownian motion in a subdiffusive Scott-Blair fluid
\begin{equation}\label{eq:Eq77}
\left( \frac{\text{6}\pi R \mu_{\alpha}}{m} \right)^{\frac{\text{2}}{\text{2}-\alpha}} \frac{m \left\langle \Delta r^{\text{2}}(t) \right\rangle}{\text{2}N K_B T} = \left( \frac{\mu_{\alpha}}{m_R} \right)^{\frac{\text{2}}{\text{2}-\alpha}} m_R J(t)
\end{equation}
as a function of the dimensionless time $\Big( ${\large $\frac{\mu_{\alpha}}{m_R}$}$t^{\text{2}-\alpha} \Big)${\large $^{\frac{\text{1}}{\text{2}-\alpha}}$} with $m_R=$ {\large $\frac{m}{\text{6}\pi R}$} for various values of the fractional exponent $\alpha \in \mathbb{R}^+$.

For the limiting case where $\alpha=$ 1, the Scott-Blair subdiffusive fluid becomes a Newtonian viscous fluid with $\mu_{\alpha}=\mu_{\text{1}}=\eta$ and Eq. \eqref{eq:Eq68} reduces to
\begin{equation}\label{eq:Eq78}
\left\langle \Delta r^{\text{2}}(t) \right\rangle = \frac{\text{2}N K_B T}{m} t^{\text{2}} E_{\text{1, 3}}\left( -\frac{t}{\tau}\right)
\end{equation}
with $\tau=$ {\large $\frac{m}{\text{6}\pi R \eta}$}. By virtue of the recurrence relation of the two-parameter Mittag--Leffler function \citep{Erdelyi1953, GorenfloKilbasMainardiRogosin2014},
\begin{equation}\label{eq:Eq79}
E_{\alpha \text{,} \beta}(z)=\frac{\text{1}}{z}E_{\alpha \text{,} \beta-\alpha}(z)-\frac{\text{1}}{z\Gamma(\beta-\alpha)}
\end{equation}
Eq. \eqref{eq:Eq78} simplifies to
\begin{equation}\label{eq:Eq80}
\left\langle \Delta r^{\text{2}}(t) \right\rangle = \frac{\text{2}N K_B T}{m}t \tau \left[ \text{1}-E_{\text{1, 2}}\left( -\frac{t}{\tau}\right) \right]
\end{equation}
\begin{figure}[b!]
\centering
\includegraphics[width=0.95\linewidth, angle=0]{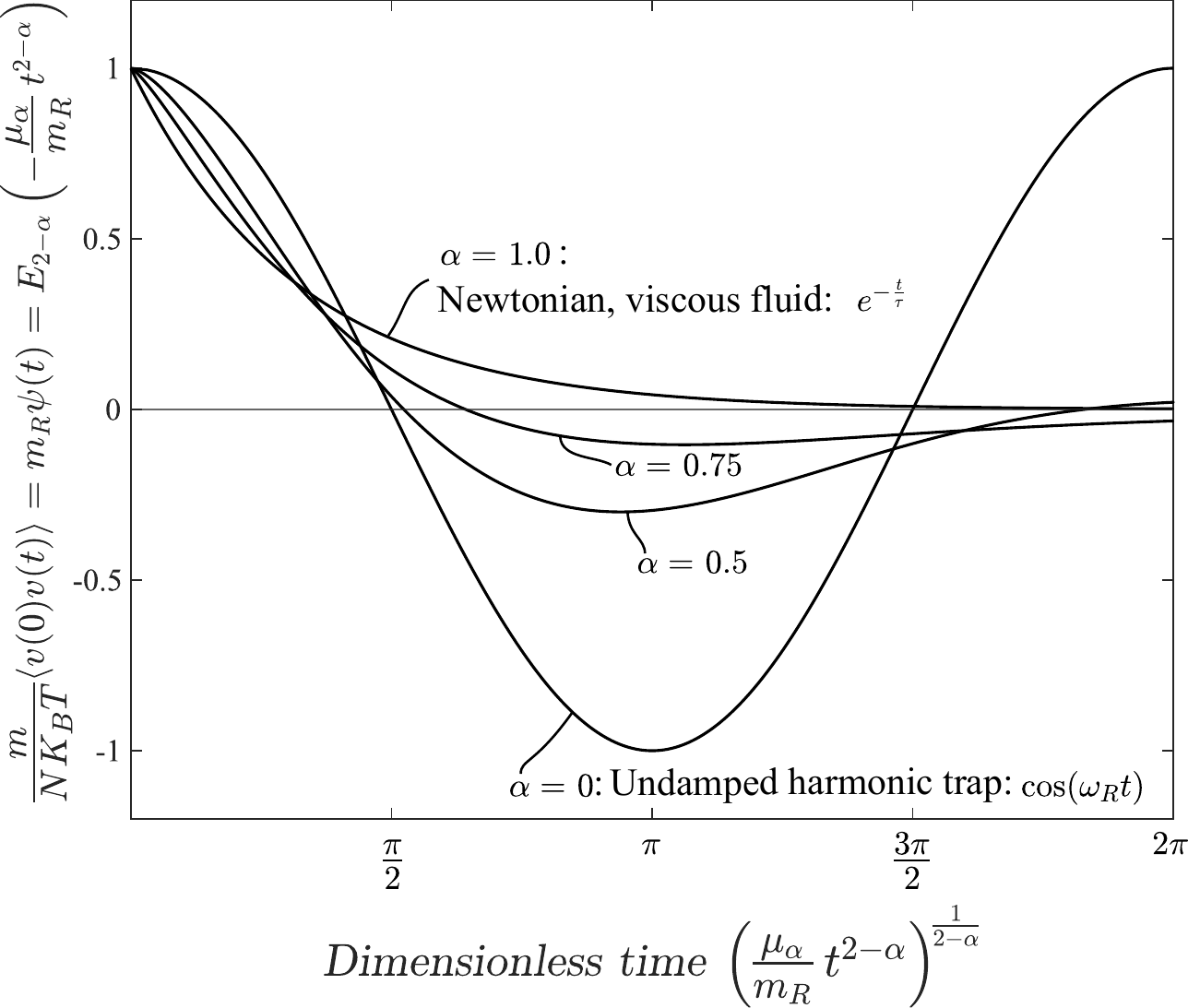}
\caption{Normalized velocity autocorrelation function of Brownian particles suspended in a fractional subdiffusive Scott-Blair fluid with material constant $\mu_{\alpha}$ with units $\left[\text{M}\right]\left[\text{L}\right]^{-\text{1}}\left[\text{T}\right]^{\alpha-\text{2}}$ for various values of the fractional exponent, 0 $\leq \alpha \leq$ 1, as a function of the dimensionless time  $\Big(${\large $\frac{\mu_{\alpha}}{m_R}$}$t^{\text{2}-\alpha} \Big)${\large $^{\frac{\text{1}}{\text{2}-\alpha}}$} where $m_R=$ {\large $\frac{m}{\text{6}\pi R}$} and $\tau=$ {\large $\frac{m}{\text{6}\pi R \eta}$}.}
\label{fig:Fig09}
\end{figure}
By using the identity, $E_{\text{1, 2}}${\large $\left( -\frac{t}{\tau} \right)$} $=$ {\large $\frac{\tau}{t}$}$\Big( \text{1}-e${\large $^{-\frac{t}{\tau}}$}$\Big)$ together with $\tau=$ {\large $\frac{m}{\text{6}\pi R \eta}$}, Eq. \eqref{eq:Eq80} further simplifies to Eq. \eqref{eq:Eq03}. Figure \ref{fig:Fig08} reveals that the mean squared displacement curves of Brownian particles suspended in a subdiffusive material $(\text{0} < \alpha \leq \text{1})$ follow the curve for Brownian motion in a Newtonian viscous fluid until time $t \approx \text{3}\tau$ and subsequently reverse their curvature to follow the power law {\large $\left(\frac{t}{\tau}\right)^{\alpha}$} \cite{PalmerXuWirtz1998, XuViasnoffWirtz1998, GislerWeitz1999}.

The complex dynamic fluidity, $\phi(s)=$ {\large $\frac{s}{\mathcal{G}(s)}$}, of the springpot--inerter parallel connection derives directly from Eq. \eqref{eq:Eq71}
\begin{equation}\label{eq:Eq81}
\phi(s)=\frac{s}{\mathcal{G}(s)}= \frac{s}{\mu_{\alpha}s^{\alpha}+m_Rs^{\text{2}}}= \frac{\text{1}}{m_R}\, \frac{s^{\text{1}-\alpha}}{s^{\text{2}-\alpha}+\frac{\mu_{\alpha}}{m_R}} \text{,} \enskip \text{0} \leq \alpha \leq \text{1}
\end{equation}
Inverse Laplace transform of Eq. \eqref{eq:Eq81} is evaluated with the convolution integral given by Eq. \eqref{eq:Eq73} where 
\begin{align}\label{eq:Eq82}
f(t)=\mathcal{L}^{-\text{1}} \left\lbrace \frac{\text{1}}{m_R} s^{\text{1}-\alpha} \right\rbrace = & \frac{\text{1}}{m_R} \frac{\mathrm{d}^{\text{1}-\alpha}\delta(t-\text{0})}{\mathrm{d}t^{\text{1}-\alpha}}= \\ \nonumber
& \frac{\text{1}}{m_R} \frac{\text{1}}{\Gamma(-\text{1}+\alpha)} \frac{\text{1}}{t^{\text{2}-\alpha}}
\end{align}
and $h(t)$ is given by Eq. \eqref{eq:Eq75}. Substitution of the results of Eqs. \eqref{eq:Eq82} and \eqref{eq:Eq75} into the convolution given by Eq. \eqref{eq:Eq73}, the impulse strain-rate response function of the springpot--inerter parallel connection is merely the fractional derivative of order $\text{1}-\alpha$ of the Rabotnov function given by Eq. \eqref{eq:Eq75}
\begin{align}\label{eq:Eq83}
\psi(t)= & \mathcal{L}^{-\text{1}} \left\lbrace \phi(s) \right\rbrace = \frac{\text{1}}{m_R}  \frac{\text{1}}{\Gamma(-\text{1}+\alpha)} \times \\ \nonumber
& \int_{\text{0}}^{t} \frac{\text{1}}{(t-\xi)^{\text{2}-\alpha}} \xi^{\text{1}-\alpha} E_{\text{2}-\alpha\text{, }\text{2}-\alpha}\left( - \frac{\mu_{\alpha}}{m_R} \xi^{\text{2}-\alpha} \right) \mathrm{d}\xi = \\ \nonumber
& \frac{\text{1}}{m_R}  E_{\text{2}-\alpha\text{, 1}}\left( - \frac{\mu_{\alpha}}{m_R} t^{\text{2}-\alpha} \right)
\end{align}

By employing the correspondence principle for Brownian motion, the velocity autocorrelation function of Brownian particles suspended in a subdiffusive Scott-Blair fluid with material constant $\mu_{\alpha}$ is given by Eq. \eqref{eq:Eq34} where now $\psi(t)$ is offered by Eq. \eqref{eq:Eq83}. This result is in agreement with the result for the velocity autocorrelation function presented by \citet{Lutz2001}. Figure \ref{fig:Fig09} plots the normalized velocity autocorrelation function as expressed by Eq. \eqref{eq:Eq56} for Brownian motion in a subdiffusive Scott-Blair fluid as a function of the dimensionless time $\Big( ${\large $\frac{\mu_{\alpha}}{m_R}$}$t^{\text{2}-\alpha} \Big)${\large $^{\frac{\text{1}}{\text{2}-\alpha}}$} with $m_R=$ {\large $\frac{m}{\text{6}\pi R}$} for various values of the fractional exponent $\alpha \in \mathbb{R}^+$. For the limiting case where $\alpha=$ 1, the Scott-Blair subdiffusive fluid becomes a Newtonian viscous fluid with $\mu_{\alpha}=\mu_{\text{1}}=\eta$ and Eq. \eqref{eq:Eq83} contracts to $\psi(t)=$ {\large $\frac{\text{1}}{m_R}$}$E_{\text{1}}\Big(-${\large $\frac{t}{\tau}$}$\Big)=$  {\large $\frac{\text{1}}{m_R}$}$e${\large $^{-\frac{t}{\tau}}$}, therefore Eq. \eqref{eq:Eq56} yields the result for the velocity autocorrelation function given by Eq. \eqref{eq:Eq33}. For the other limiting case where $\alpha=$ 0, the Scott-Blair fluid becomes a Hookean solid with $\mu_{\alpha}=\mu_{\text{0}}=G$ and Eq. \eqref{eq:Eq83} contracts to $\psi(t)=$ {\large $\frac{\text{1}}{m_R}$}$E_{\text{2}}(-\omega_R^{\text{2}}t^{\text{2}})=$ {\large $\frac{\text{1}}{m_R}$}$\cos(\omega_R t)$ where $\omega_R=$ {\large $\sqrt{\frac{G}{m_R}}$}. The $\cos(\omega_R t)$ periodic response is in agreement with the results for Brownian motion in an undamped harmonic potential $(\eta=\text{0})$ given by Eq. \eqref{eq:Eq35} after setting {\large $\frac{\text{1}}{\tau}$} $=$ {\large $\frac{\text{6}\pi R \eta}{m}$} $=$ {\large $\frac{\eta}{m_R}$} $=$ 0.

\section{Summary}
This paper builds upon past theoretical and experimental work on Brownian motion and microrheology and uncovers that for all time-scales the mean squared displacement of Brownian microspheres with mass $m$ and radius $R$ suspended in any linear, isotropic viscoelastic material is identical to the creep compliance of a linear mechanical network that is a parallel connection of the linear viscoelastic material with an inerter with distributed inertance $m_R=$ {\large $\frac{m}{\text{6}\pi R}$}. The synthesis of this mechanical network simplifies appreciably the calculation of the mean squared displacement and the velocity autocorrelation function of Brownian particles suspended in viscoelastic materials where inertia effects are non-negligible at longer time-scales as is the case of a Maxwell fluid. The viscous--viscoelastic correspondence principle established in this paper after introducing the concept of the inerter is equivalent to the viscous--viscoelastic analogy suggested by \citet{MasonWeitz1995}; while it extends the work of \citet{PalmerXuWirtz1998} and \citet{XuViasnoffWirtz1998} for all fluid-like and solid-like viscoelastic materials at all time-scales.

The proposed correspondence principle was employed to calculate the mean squared displacement and the velocity autocorrelation function of Brownian particles suspended in a fractional subdiffusive Scott-Blair fluid and concludes that for times $t>\text{3}\tau=$ {\large $\frac{m}{\text{2}\pi R \eta}$}, the mean squared displacement reverses its curvature and follows the power law {\large $\left(\frac{t}{\tau}\right)^{\alpha}$} as was shown in the experiments reported in the literature \cite{PalmerXuWirtz1998, XuViasnoffWirtz1998, GislerWeitz1999}. The study concludes that for Brownian particles suspended in a subdiffusive Scott-Blair fluid $(\text{0} \leq \alpha \leq \text{1})$, the mean squared displacement is the fractional integral of order $\text{1}+\alpha$ of the Rabotnov function {\large $\varepsilon$}$_{\text{1}-\alpha}\Big(-${\large $\frac{\mu_{\alpha}}{m_R}$}, $t\Big)=t^{\text{1}-\alpha}E_{\text{2}-\alpha\text{, 2}-\alpha}\Big(-${\large $\frac{\mu_{\alpha}}{m_R}$}$t^{\text{2}-\alpha}\Big)$ \citep{Rabotnov1980}; whereas, the velocity autocorrelation function is the fractional derivative of order $\text{1}-\alpha$ of the same Rabotnov function, {\large $\varepsilon$}$_{\text{1}-\alpha}\Big(-${\large $\frac{\mu_{\alpha}}{m_R}$}, $t\Big)$.

\bibliographystyle{apsrev4-2.bst}
\bibliography{References} 
\end{document}